\documentclass[aps,pre,twocolumn,longbibliography]{revtex4-2}

\usepackage[pdftex]{graphicx}
\usepackage{xcolor}
\usepackage{gensymb}
\usepackage{bm}

\begin{document}
\title{Predicting elastic and plastic properties of small iron polycrystals by machine learning}
\author{Marcin Mi{\'n}kowski(\url{marcin.minkowski@tuni.fi})}
\author{Lasse Laurson}
\affiliation{Computational Physics Laboratory, Tampere University, P.O. Box 692, FI-33014 Tampere, Finland}

\begin{abstract}
Deformation of crystalline materials is an interesting example of complex system behaviour. Small samples typically exhibit a stochastic-like, irregular response to externally applied stresses, manifested as significant sample-to-sample variation in their mechanical properties. In this work we study the predictability of the sample-dependent shear moduli and yield stresses of a large set of small cube-shaped iron polycrystals generated by Voronoi tessellation, by combining molecular dynamics simulations and machine learning. Training a convolutional neural network to infer the mapping between the initial polycrystalline structure of the samples and features of the ensuing stress-strain curves reveals that the shear modulus can be predicted better than the yield stress. We discuss our results in the context of the sensitivity of the system's response to small perturbations of its initial state.
\end{abstract}
\maketitle
\section{Introduction}
Crystalline materials studied in experiments are almost never perfect monocrystalline structures. Most often they contain lattice defects and are usually polycrystals, i.e., they are composed of several grains of different lattice orientations separated by grain boundaries, which play a crucial role in determining the mechanical properties of the sample \cite{kheradmand2010investigation}. During their deformation the complexity of the dynamics of the polycrystal on the microscopic scale makes predicting the mechanical response of a single sample based on its initial state (microstructure) challenging. Moreover, crystal plasticity exhibits size effects implying that smaller systems are stronger (the stress required to reach a given strain is higher) and their mechanical response to the externally applied stresses tends to be irregular and is characterized by a significant sample-to-sample variation \cite{uchic2009plasticity,dimiduk2005size}. The latter features originate from the sample-dependent microstructure of small polycrystals, implying that predicting their mechanical response is likely to be particularly challenging.

In recent years a huge progress in development and application of machine learning (ML) techniques in many fields of science has been observed \cite{janiesch2021machine,carleo2019machine,shinde2018review,ongsulee2017artificial,schmidt2019recent,rajendra2022advancement}. In material science it has led to emergence of methods able to identify and characterize samples \cite{stoll2021machine,chan2020machine,baskaran2020adaptive}, to design novel materials with desired properties \cite{vasudevan2021machine,durodola2022machine,moosavi2020role,wei2019machine}, and to establish relations between the structure and the properties of the material \cite{dai2020method,fu2022review,chibani2020machine,jung2019efficient}. A related research problem, relevant for the present study, is predicting the mechanical response of a sample of material during its deformation \cite{salmenjoki2018machine,sarvilahti2020machine,minkowski2022machine}. The general problem statement can be formulated as follows: Given some description of the initial state (microstructure) of the sample, with what accuracy can its mechanical response be predicted?

The accuracy of the prediction of the given ML algorithm can be expressed quantitatively for instance by the coefficient of determination $r^2$. If the system studied is governed by deterministic equations of motion, in principle it should be possible to train an algorithm to represent its dynamics perfectly, which would result in perfect predictability score $r^2=1$. In practice, however, this does not usually happen. The dynamics of many complex systems is to some degree chaotic, or as in the case of dislocation dynamics, exhibits critical behaviour \cite{ovaska2015quenched,papanikolaou2017obstacles,zapperi2001depinning,chan2010plasticity,alava2014crackling}. This implies that the time evolution of a complex system such as a small plastically deforming crystal may be sensitive to small perturbation of its initial conditions. In other words, perturbing slightly the initial state of the system can lead to significant differences in its subsequent dynamics. This limits the extent to which the time evolution of such systems can be predicted (e.g., via ML algorithms) because the full information of the initial state, which on the atomic scale includes positions and velocities of all the atoms, is usually not available due to the finite precision of any experimental observations or coarse-grained numerical representations of the data. Moreover, due to finite decimal precision numerical simulations are never perfectly accurate either, something that may further amplify the differences caused by small perturbations of the initial state. This study concerns computer simulations only, but as discussed above, the lack of full description of the initial state exists also in experiments, where any characterization of the initial microstructure (using various imaging techniques) has a finite precision.

Polycrystals have been studied by ML in several publications \cite{shu2022grain,vieira2021machine,hestroffer2023graph,dai2021graph,karimi2023prediction}, where experimental data and finite element simulations were used to produce the training data. In contrast, in this work, we study predictability of the deformation process of cube-shaped iron nanopolycrystals by combining strain-controlled molecular dynamics (MD) simulations with ML methods. Employing MD simulations allows to take into account atomistic details of the structure, which is especially important in polycrystals due to existence of grain boundaries. We generate a large set of polycrystals with various shapes and sizes of the grains and use it to train a convolutional neural network (CNN) to infer the link between the initial microstructure and features of the stress-strain curve. Our study focuses on a specific morphology of polycrystals, since for their generation we use Voronoi tessellation. We show that the key elastic and plastic properties characterizing the response of the system to applied shear stresses, namely shear modulus and yield stress, exhibit different degree of predictability, measured here by the coefficient of determination $r^2$. As the descriptors for CNN we use fields describing the local properties of the polycrystals on the atomic level. The degree of predictability we find for these quantities is then discussed in the context of the sensitivity of the system to small perturbations of the initial conditions. We propose that sensitivity is an important factor giving rise to fundamental limits to predictability of evolution of complex systems such as deformation predictability.

\begin{figure*}
	\centering
	\includegraphics[scale=0.5]{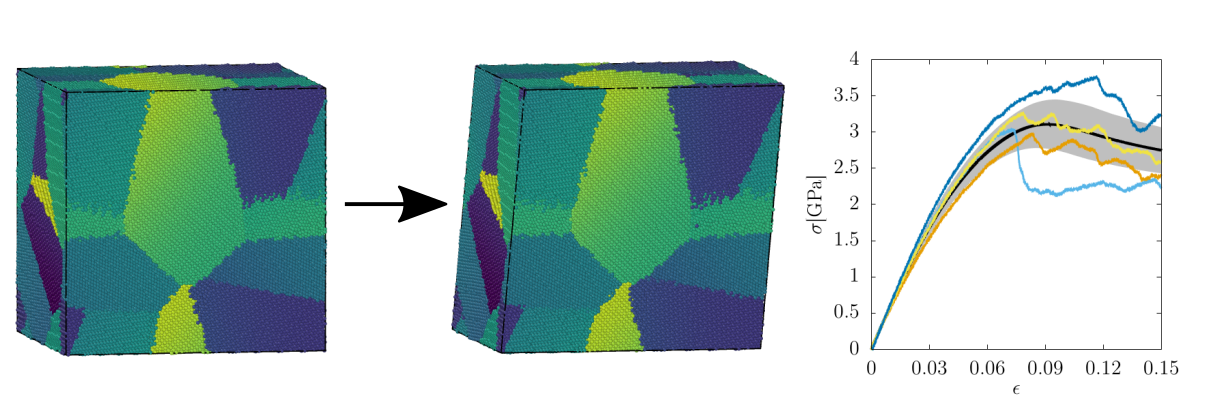}
	\caption{Schematic of the model studied in the work. The polycrystalline sample is first equilibrated at 300 K (left), after which it is shear deformed by MD simulations with a constant strain rate (middle). During the simulation the instantaneous shear stress $\sigma$ is measured as a function of the strain $\epsilon$, resulting in a unique stress-strain curve $\sigma(\epsilon)$ for each sample. Repeating the simulation several times for different initial polycrystal structures results in an ensemble of stress-strain curves with a mean shown as the black line and a standard deviation shown with gray (right).}
	\label{model}
\end{figure*}
\begin{figure*}
	\centering
	\includegraphics[scale=0.35]{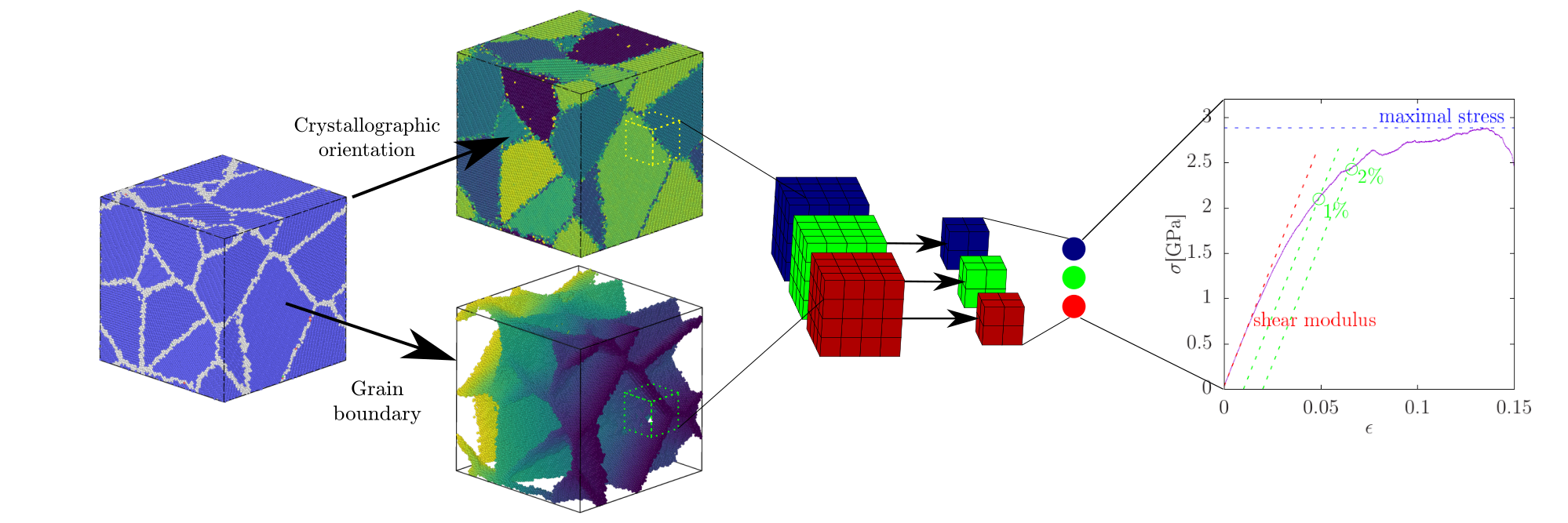}
	\caption{Schematic of the CNN used for predicting the shear modulus and yield stress of the polycrystal. Three-dimensional arrays representing the local crystallographic orientation and the grain boundaries of the polycrystal are fed into three-dimensional convolutional layers, where they are processed by the convolutional filters. Different colours represent different filters. The size of the arrays is subsequently reduced by the maximum pooling layer. The procedure is repeated until the arrays are of the size $1\times1\times1$. They are then concatenated and fed into the fully connected neural network, which gives the final output, i.e., either shear modulus or the yield stress according to one of the definitions indicated in the right panel. The percentage values refer to the offset yield method.}
	\label{cnn}
\end{figure*}
\section{Results}
\label{section:results}
\subsection{Deformation of iron polycrystals}
First a set of 4000 cube-shaped iron polycrystalline initial configurations is generated with Voronoi tessellation. The sample size is $20\times20\times20$ nm$^{3}$ and all samples contain 8 nanograins of the average size 10 nm with randomly chosen positions of the seeds in the Voronoi tessellation and Euler angles specifying the crystallographic lattice orientation. Even though both the size of the nanocrystal and the number of nanograins are fixed, the individual grains are of different shapes and sizes, and therefore, the volume fraction of the boundary between them also varies from sample to sample. The lattice structure is bcc and the lattice constant is chosen to be 0.287 nm. Each configuration contains around 677000 atoms. More details about generation of iron polycrystals can be found in the Methods section \ref{subsection:methodsPolycrystals}.

After the set of initial polycrystalline configurations is generated, their energy is first minimized by adjusting atom coordinates and afterwards they are equilibrated at 300 K. During those phases the initially sharp grain boundaries, as generated by Voronoi tessellation, transform slightly by local atomic rearrangement. Subsequently the MD simulations of shear deformation are carried out for each of them by Large-scale Atomic/Molecular Massively Parallel Simulator (LAMMPS) \cite{LAMMPS} (see Methods section \ref{subsection:methodsMD} for more details), allowing to obtain the sample-specific stress-strain curve for each sample. The model is shown schematically together with a few corresponding example stress-strain curves in Fig. \ref{model}. As can be observed there, the curves exhibit a large sample-to-sample variability. While the elastic part of different curves is similar (but not the same), increasing linearly with a certain slope whose magnitude varies from sample to sample, in the plastic regime, which typically starts around the strain value of 0.09, there are large differences in the stress response and the curves exhibit fluctuating character with many stress drops. One can also observe that different curves exhibit qualitatively different behaviour. Some of them have a large stress drop at some value of strain, while others after reaching the yield remain relatively flat. As a consequence, the yield stress exhibits a much larger variability than the shear modulus.

For each configuration the shear modulus and yield stresses (using different definitions, see below) are extracted from the corresponding stress-strain curves obtained during the simulation. The shear modulus is taken as the slope of the linear function fitted with the least square method to the stress-strain curve in the range of strain from 0 to 0.01, in which the system is still in the elastic regime. For extracting the yield stress, on the other hand, several different definitions are applied, which are discussed below.

As the descriptors input fields representing the local lattice orientation and density of atoms at the grain boundary are extracted with different resolutions from the equilibrated configurations. Along with the output values of the shear modulus and yield stress extracted from the stress-strain curves they are subsequently used to train CNNs. The schematic of the CNN is shown in Fig. \ref{cnn} (see Methods section \ref{subsection:methodsDescriptors} for details on descriptors and Methods section \ref{subsection:methodsCNN} for details on CNN architecture). The predictability is measured as the coefficient of determination $r^2$ as the function of the dataset size $N$, given by
\begin{equation}
r^2=1-\frac{\sum_{i=1}^{N}(y_i-f_i)^2}{\sum_{i=1}^{N}(y_i-\langle y \rangle)^2},\label{coeff_determ}
\end{equation}
where $y_i$ is the true value of shear modulus or yield stress of the sample $i$, $\langle y \rangle$ is the mean value, $f_i$ is the value predicted by the CNN, and $N$ is the total number of samples in the given set. The coefficient of determination is a commonly used metric to assess the quality of the regression analysis and can be interpreted as the proportion of the variance in the dependent (predicted) variable that is predicted from the independent variables (input) \cite{chicco2021coefficient}.
\subsection{Shear modulus}
A material is said to be in elastic regime when it returns to its original shape and size after the externally applied stress is removed. Elasticity is quantitatively characterized by a set of elastic constants, such as Young's modulus, bulk modulus or shear modulus, which indicate what amount of stress is needed to deform the sample in a certain way. Those constants can be written in the form of the elasticity tensor.

While the elastic constants of monocrystals are known for most materials, in the case of polycrystalline samples they depend on the shape and crystallographic orientation of each constituent grain \cite{sheng2012effective}. The elastic constants of those individual grains correspond to the rotational transformation of the elasticity tensor obtained for the main crystallographic axes. Moreover, in the equilibrated polycrystalline sample the crystallographic orientation may be different near the grain boundaries than within the grains, which may also influence the elastic properties of the whole material. One can thus expect that the shear modulus of the whole polycrystal can be extracted with a reasonable accuracy from the field of the crystallographic orientation varying within the sample.

\begin{figure}
	\centering
    \includegraphics[scale=0.9]{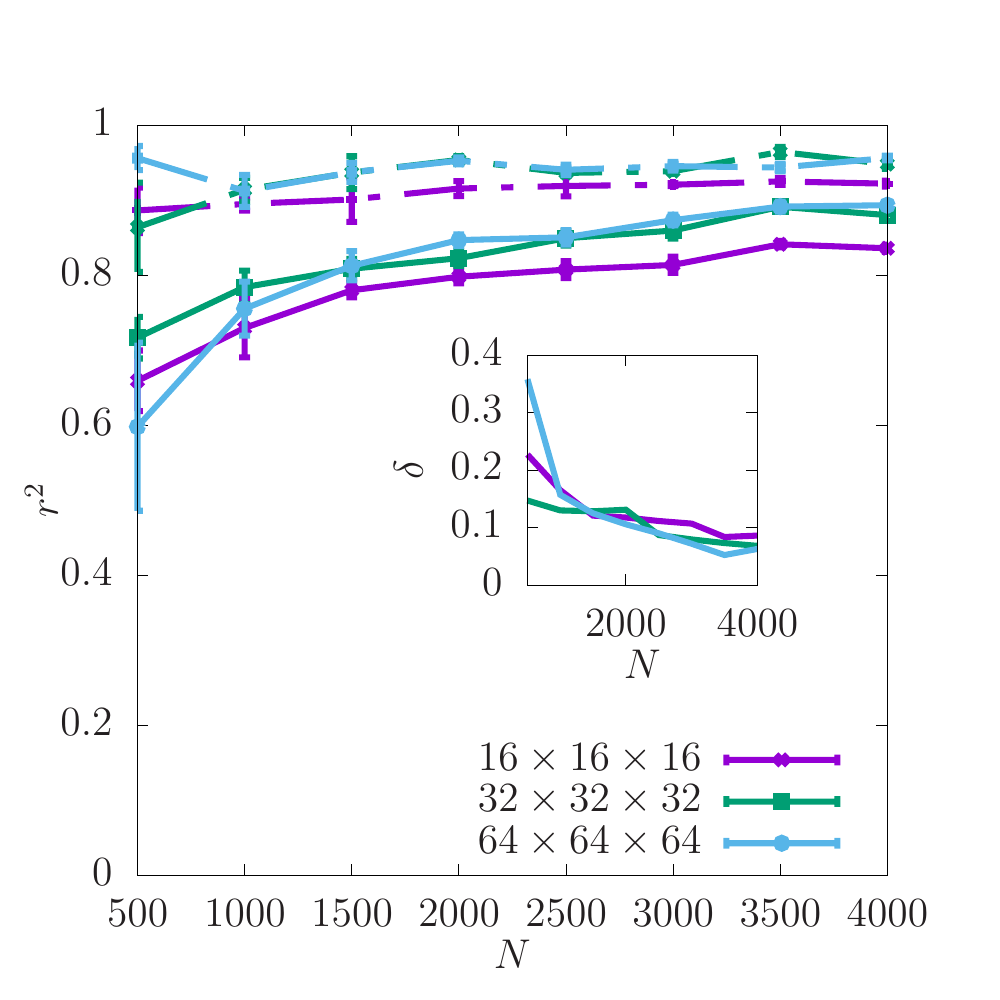}
    \caption{$r^2$ for shear modulus obtained for three different resolutions of the CNN input data as a function of $N$. The dashed lines show the values of $r^2$ for the training set, and the continuous lines for the test set. The inset shows $\delta$ as a function of $N$. The errorbars are standard errors of the mean (SEM).}
    \label{shear}
\end{figure}

\begin{figure*}
    \centering
    \includegraphics[scale=0.55]{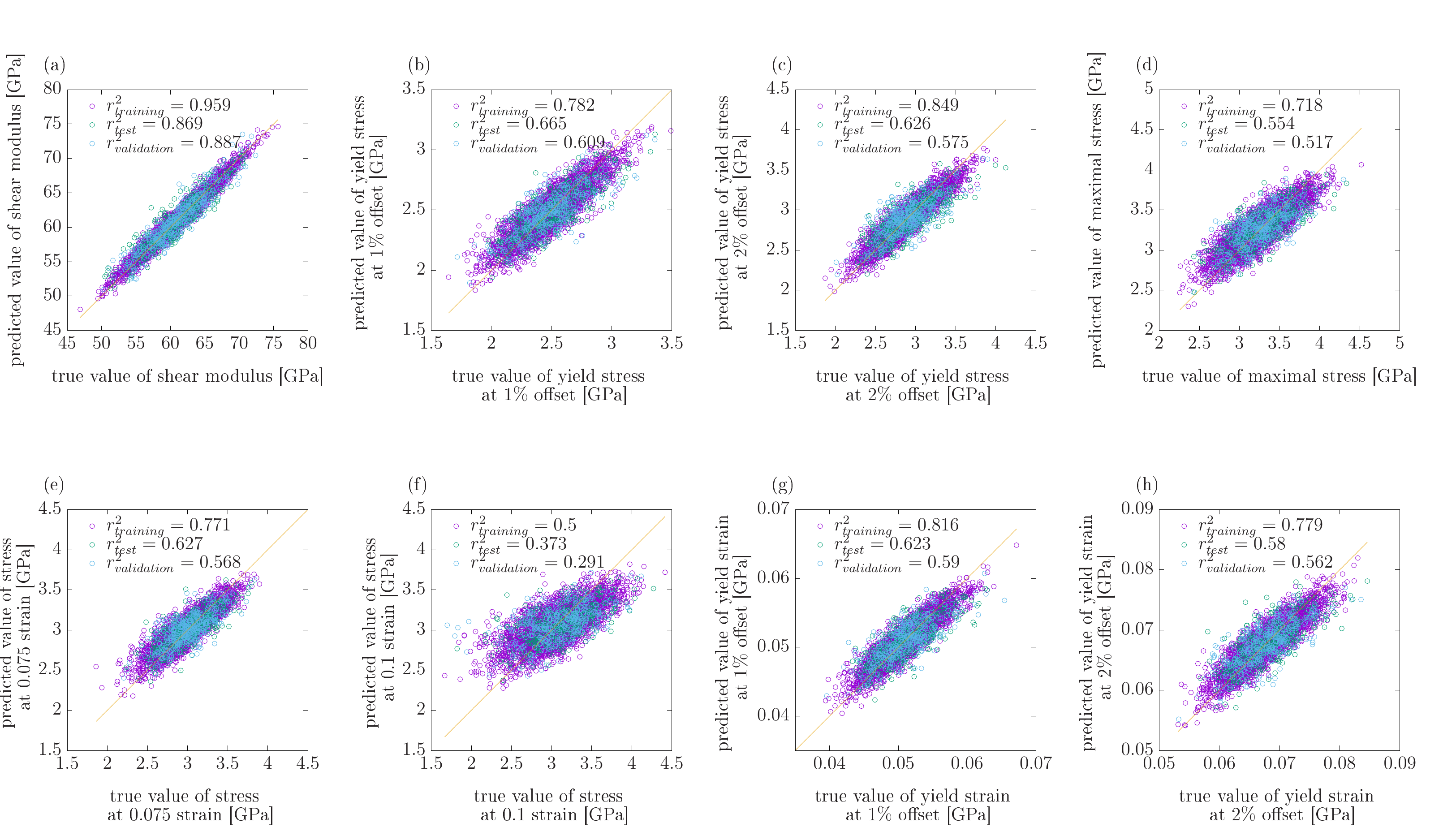}
    \caption{Scatter plots of true values of the quantities studied in the work against the corresponding values predicted by the CNN for one single seed with the resolution of descriptors $32\times32\times32$.}
    \label{scatter_plots}
\end{figure*}

\begin{figure}
	\centering
    \includegraphics[scale=0.9]{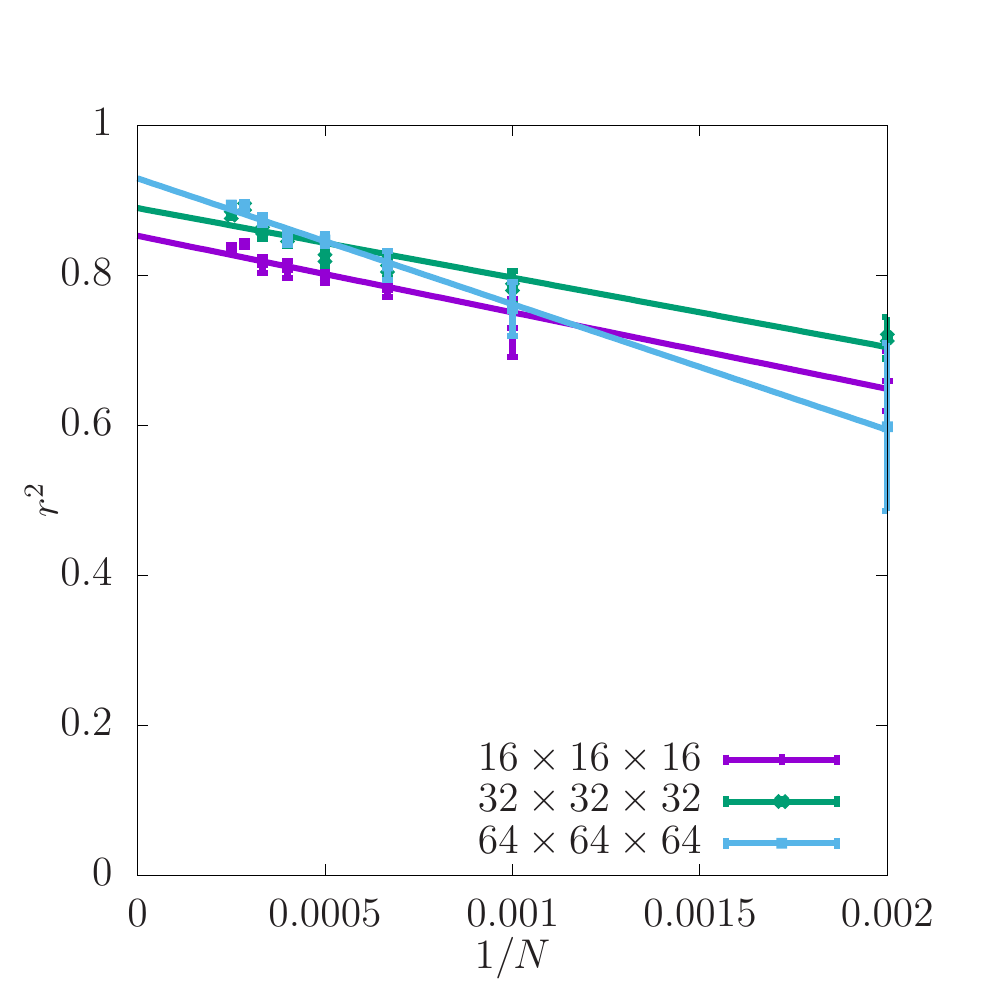}
    \caption{Asymptotic behaviour of $r^2$ of the test set for shear modulus obtained for three different resolutions of the CNN input data as a function of $1/N$. The errorbars are standard errors of the mean (SEM).}
    \label{shear_asymptotic}
\end{figure}

In Fig. \ref{shear} the coefficient of determination $r^2$ is shown for the shear modulus as a function of the dataset size $N$ used as the input for training the CNN. The predictability is already good even for the smallest values of $N$. Adding more configurations increases it further and reduces the training-test set gap $\delta$ as seen in the insets. Moreover, it can be seen that increasing the resolution of the input data also improves the predictability. While the difference in $r^2$ between the resolutions of $16\times16\times16$ and $32\times32\times32$ is quite significant, the results do not improve much more when the resolution is increased further up to $64\times64\times64$. In Fig. \ref{scatter_plots}a the scatter plot of the true against the predicted values of the shear modulus is shown for one of the seeds with the resolution of descriptors $32\times32\times32$. Fig. \ref{shear_asymptotic} shows $r^2$ of the test set for the shear modulus as a function of the inverse dataset size, $1/N$, with a linear function fitted to the points. It allows to estimate the asymptotic value of $r^2$ for $N\rightarrow\infty$.

\begin{figure}
    \centering
    \includegraphics[scale=0.9]{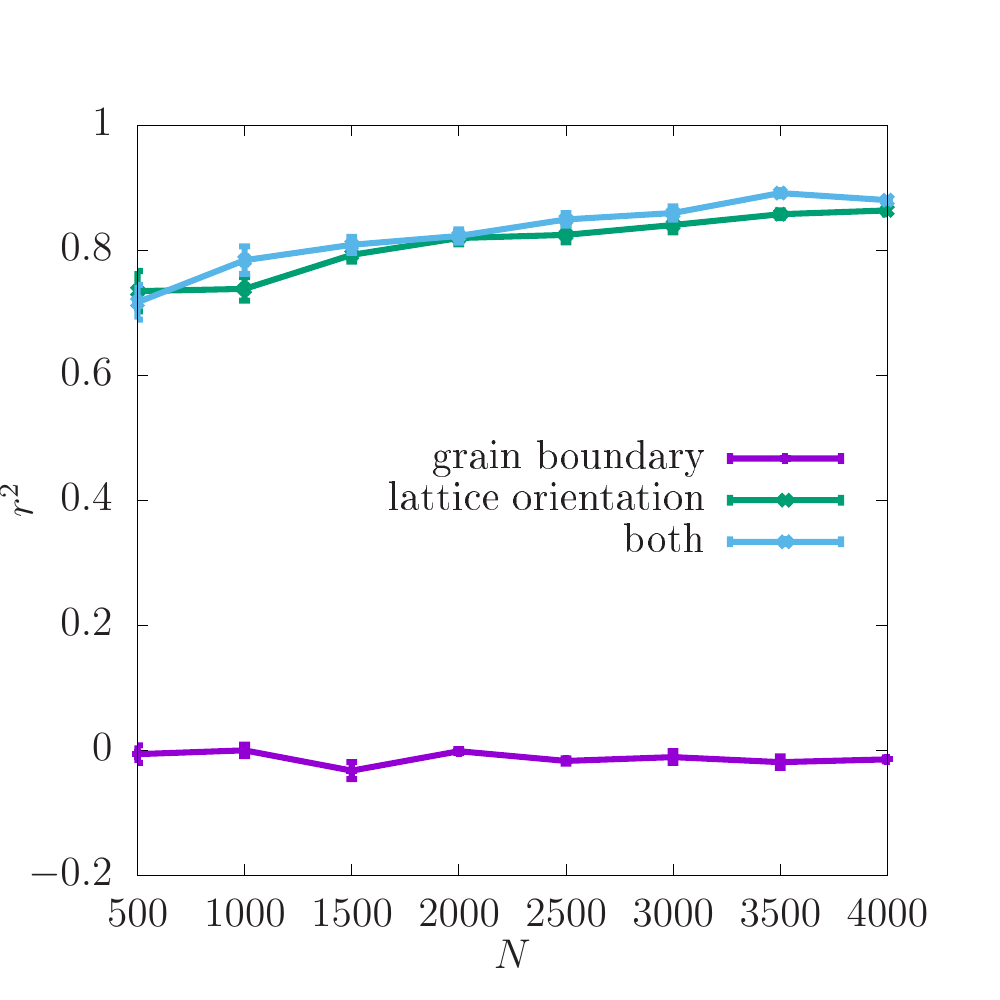}
    \caption{$r^2$ for shear modulus obtained for the test set at the resolution $32\times32\times32$ with the descriptors used separately and combined as a function of $N$. The errorbars are standard errors of the mean (SEM).}
    \label{shear_descriptors}
\end{figure}

Additionally, the values of $r^2$ resulting from training with only one of the descriptors compared to $r^2$ for both descriptors combined are shown in Fig. \ref{shear_descriptors} for the resolution $32\times32\times32$. As one can see there, the lattice orientation of the individual grains of the polycrystal is a more important descriptor for predicting the shear modulus than the grain boundary. The values of $r^2$ for the latter descriptor are in fact slightly negative, which suggests that it does not on its own provide any information about the shear modulus. However, $r^2$ for both the descriptors combined is still slightly higher than that for the lattice orientation only in almost the whole range of $N$. This suggests that the descriptor of the grain boundary may in fact contain some information relevant to the shear modulus but only in combination with the other descriptor.
\subsection{Yield point}
Crystals deform plastically when they do not return to their original shape if the external stress is removed. While in a perfect crystal 
slip, which is displacement of two atomic layers along each other, would require a large amount of stress, plasticity in real crystals is facilitated by defects. Very often the plasticity is mediated by motion of dislocations. In the polycrystals studied here no dislocations are present within the grains in the initial configuration, however, the grain boundaries with low misorientation angles can be considered as arrays of dislocations. Polycrystals will most often deform via nucleation of dislocations from the grain boundaries and by grain boundary sliding~\cite{weissmuller2011kinematics,patriarca2013slip}.

Yield of the material is the point on the stress-strain curve that indicates the transition from the elastic to the plastic behaviour. Once the sample enters the plastic regime it is deformed permanently. The yield point is fully specified by giving its two coordinates, referred to as the yield strain and yield stress. There are various ways in which the position of the yield point is precisely determined.

In metals the yield point is defined by the offset method \cite{wei2007plane}. In that approach the yield is determined as the intersection of the stress-strain curve with the line parallel to the elasticity region. The offset by which that line is shifted may vary depending on the specific material. Usually it is chosen as 0.002 (0.2\%) strain \cite{ross1999mechanics,wei2007plane}, however, that value is not particularly useful in this work because the intersection point determined with it lies within the elastic part of the stress-strain curve. It has been shown that nanocrystalline polycrystals deform more heterogeneously, and therefore, not all the grains are deformed at all by the the 0.2\% offset stress \cite{brandstetter2006micro,saada2005hall}.  Hence, instead, in this work the offset values 0.01 (1\%) and 0.02 (2\%) are chosen. The method is illustrated in the right panel of Fig. \ref{cnn}. As can be seen there the offset value for 1\% is still in the pseudo-elastic regime, which just corresponds to strain softening of the sample. However, the point determined with that method for 2\% is located just slightly behind the first abrupt change in the slope of the stress-strain curve. The yield point determined by the offset method can be interpreted as the state of the plastic deformation of the system by the value specified by the offset.

One can also be interested in the maximal stress that the sample can withstand \cite{dinkgreve2016different,moller2006yield}, which corresponds to the global maximum of the stress-strain curve. However, in the case of irregular, highly fluctuating stress-strain curves, such as those occurring in small samples, that definition might not be appropriate as a definition of the yield stress.

One can also consider the stress value at some fixed strain, which is sometimes defined as the flow stress \cite{ishikawa2005high,luo2016correlation}. The exact value of the strain should be chosen in such a way that the system has been already plastically deformed. Looking at the average stress-strain curve in Fig. \ref{model} an appropriate value of strain to choose is approximately in the range of 0.075-0.1.
\subsubsection{Yield stress}
\begin{figure}
    \centering
    \includegraphics[scale=0.9]{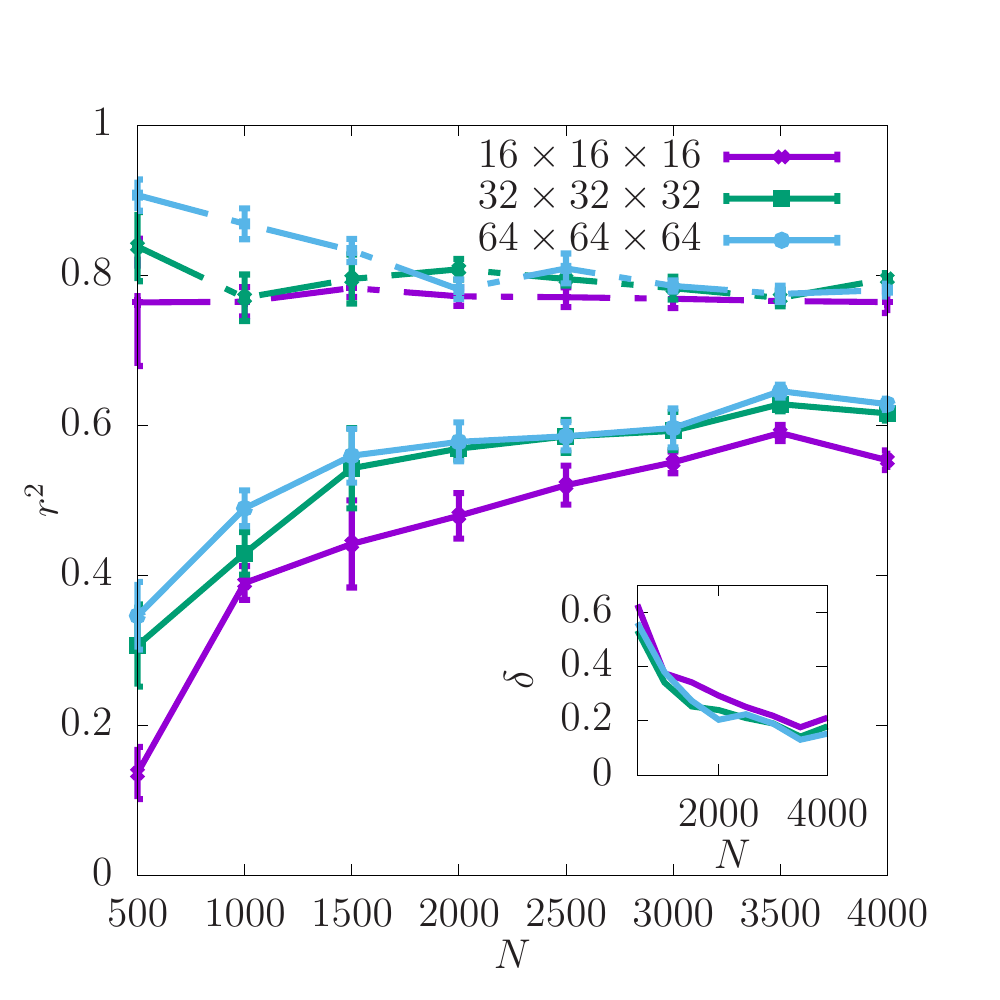}
    \caption{$r^2$ for the yield stress defined with the offset method at 2\% obtained for three different resolutions of the CNN input data as a function of $N$. The dashed lines show the values of $r^2$ for the training set, and the continuous lines for the test set. The errorbars are standard errors of the mean (SEM). The inset shows $\delta$ as a function of the dataset size.}
    \label{offsetYieldStressResolutions}
\end{figure}
\begin{figure}
    \centering
    \includegraphics[scale=0.9]{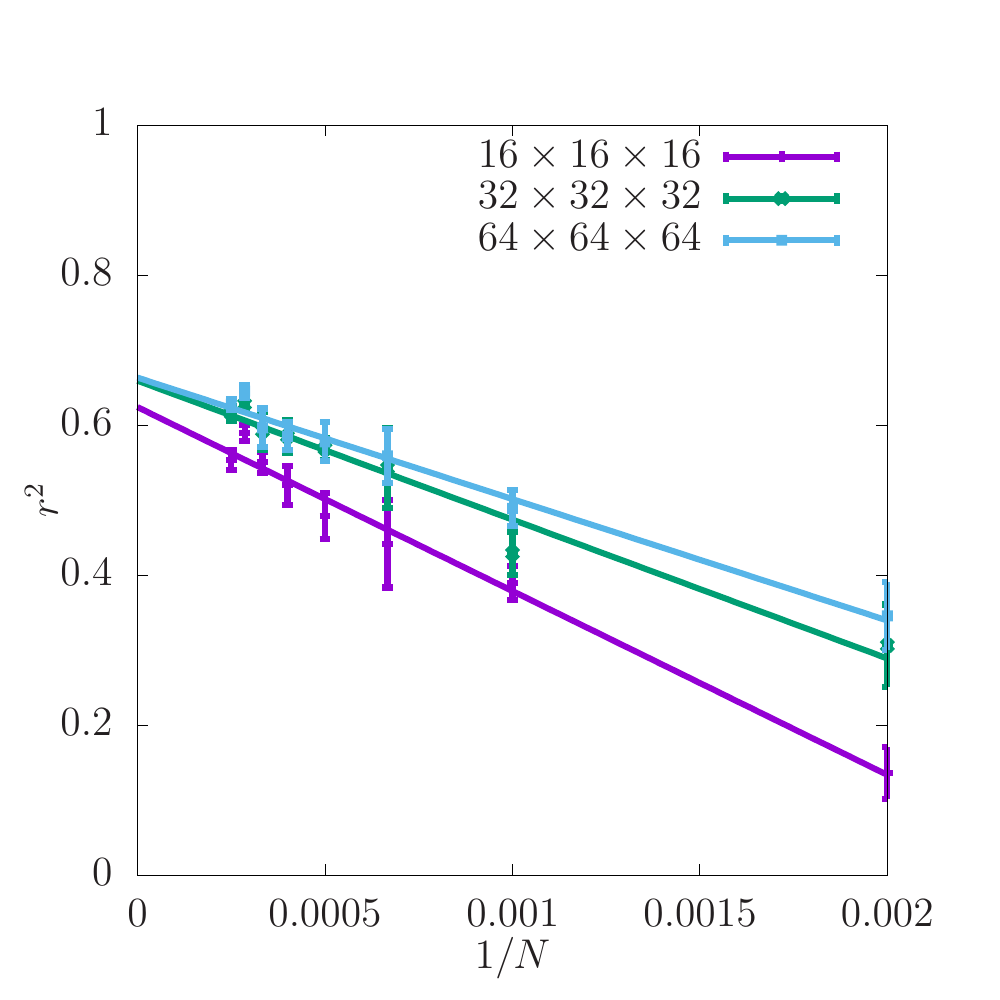}
    \caption{Asymptotic behaviour of $r^2$ of the test set for the yield stress defined with the offset method at 2\% obtained for three different resolutions of the CNN input data as a function of $1/N$. The errorbars are standard errors of the mean (SEM).}
    \label{offsetYieldStressResolutions_asymptotic}
\end{figure}
\begin{figure}
    \centering
    \includegraphics[scale=0.9]{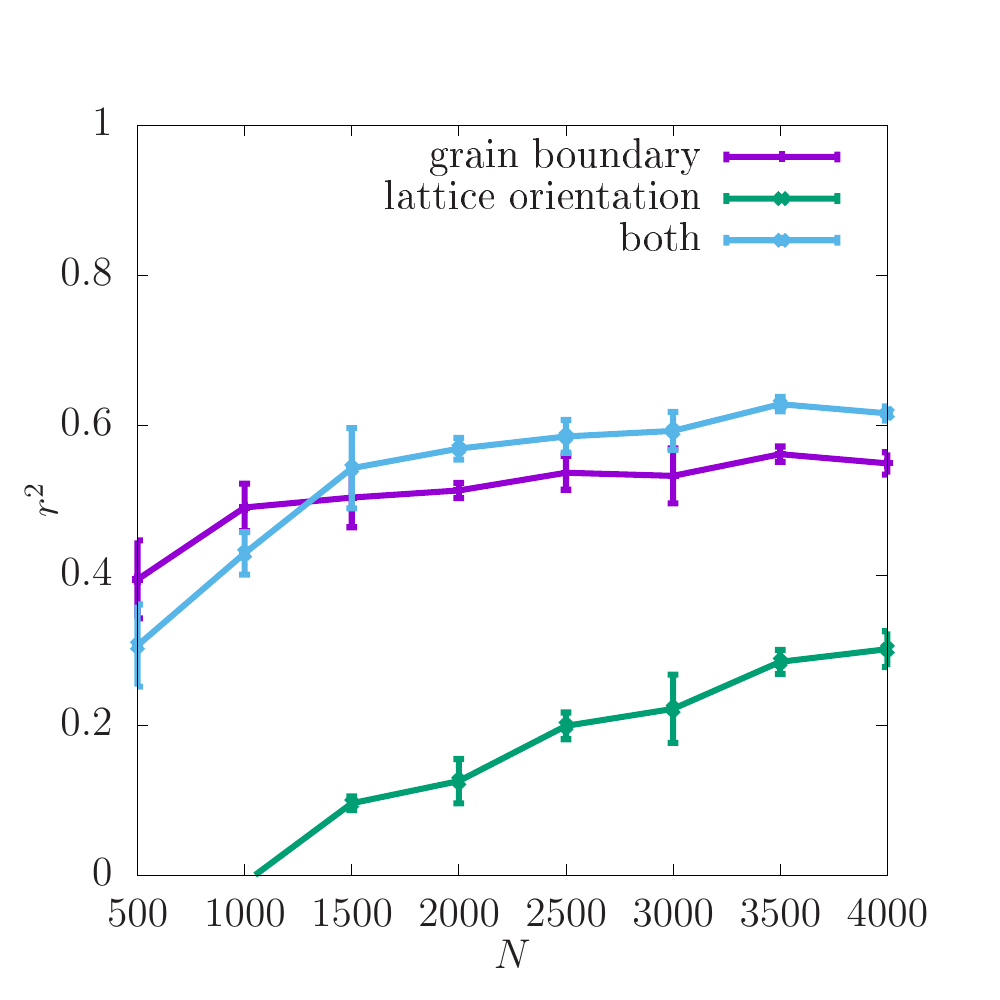}
    \caption{$r^2$ for the yield stress defined with the offset method at 2\% obtained for the test set at the resolution $32\times32\times32$ with the descriptors used separately and combined. The errorbars are standard errors of the mean (SEM).}
    \label{offsetYieldStressDescriptors}
\end{figure}
\begin{figure}
	\centering
    \includegraphics[scale=0.9]{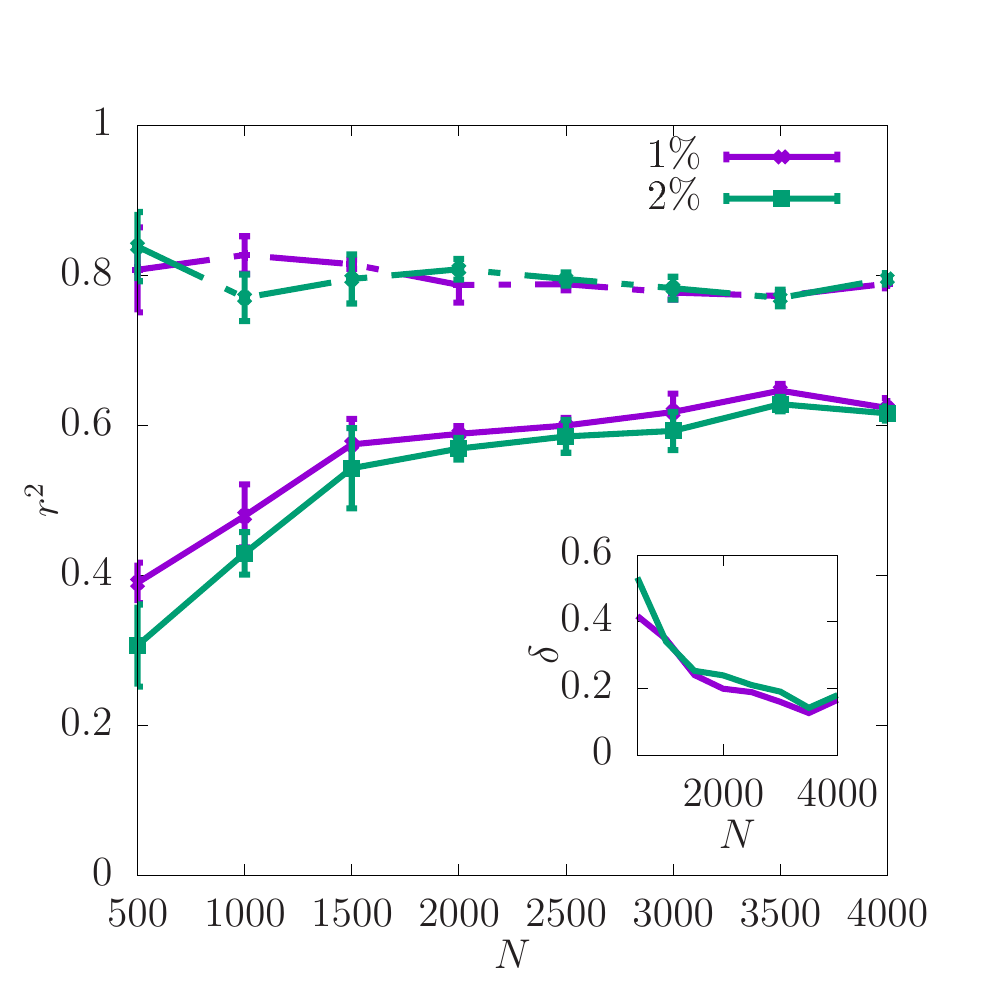}
    \caption{$r^2$ for the yield stress defined with the offset method for the resolution of $32\times32\times32$ of the CNN input data as a function of $N$. The dashed lines show the values of $r^2$ for the training set, and the continuous lines for the test set. The inset shows $\delta$ as a function of $N$. The errorbars are standard errors of the mean (SEM).}
    \label{offsetYieldStress}
\end{figure}
\begin{figure}
	\centering
    \includegraphics[scale=0.9]{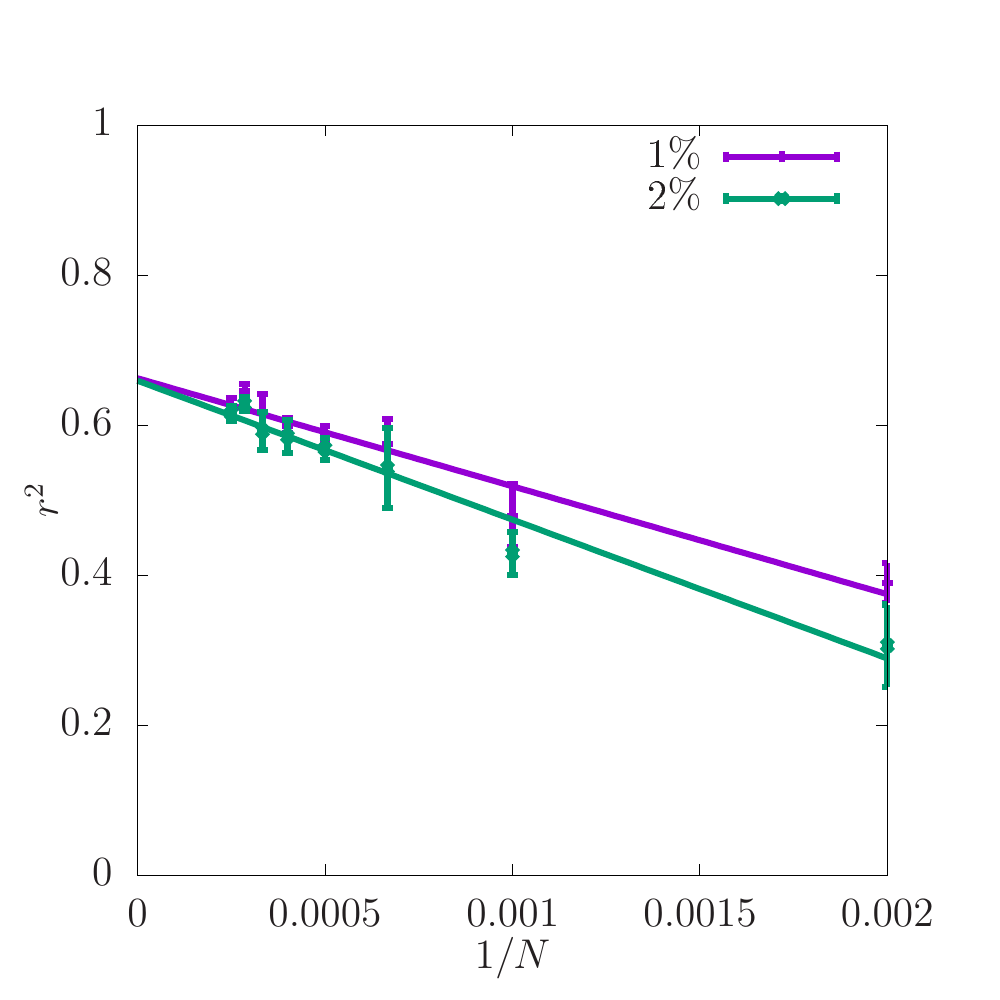}
    \caption{Asymptotic behaviour of $r^2$ of the test set for the yield stress defined with the offset method for the resolution of $32\times32\times32$ of the CNN input data as a function of $1/N$. The inset shows $\delta$ as a function of $N$. The errorbars are standard errors of the mean (SEM).}
    \label{offsetYieldStress_asymptotic}
\end{figure}
\begin{figure}
	\centering
    \includegraphics[scale=0.9]{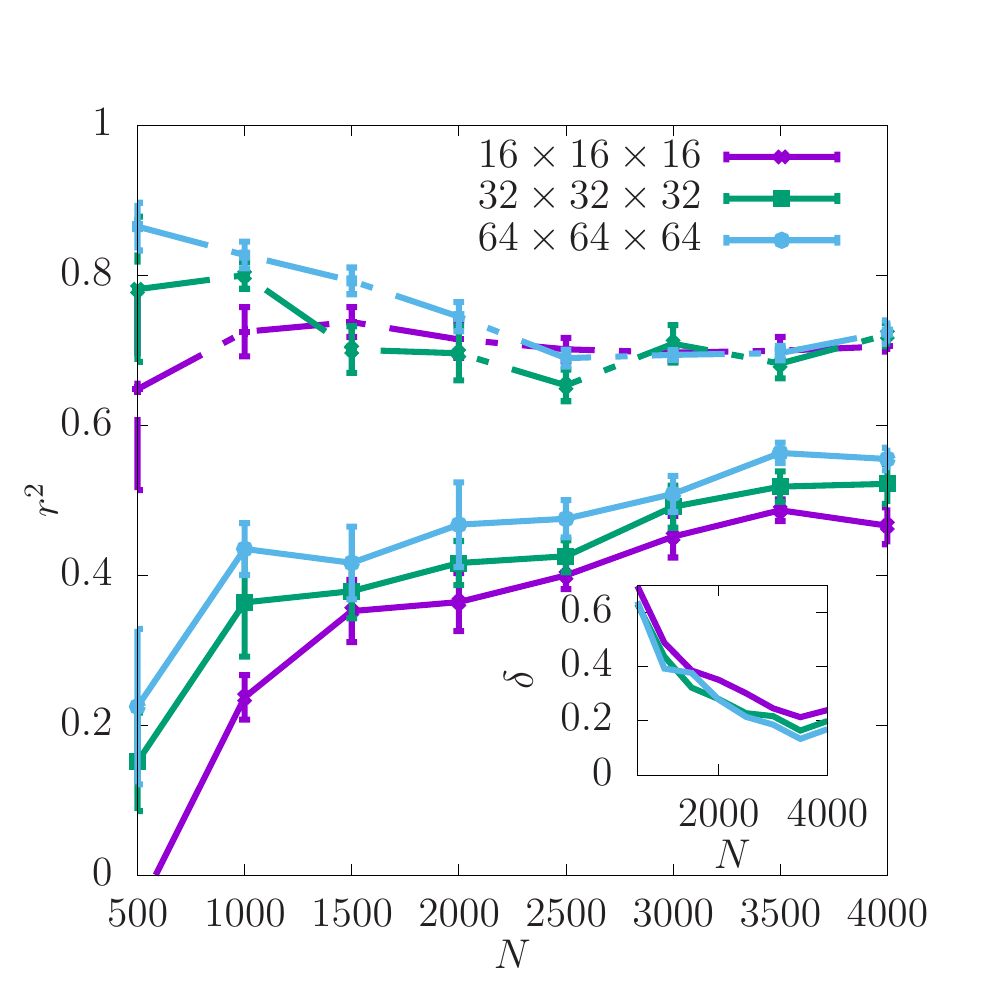}
    \caption{$r^2$ for the maximal stress value along the stress-strain curve obtained for three different resolutions of the CNN input data as a function of $N$. The dashed lines show the values of $r^2$ for the training set, and the continuous lines for the test set. The errorbars are standard errors of the mean (SEM). The inset shows $\delta$ as a function of the dataset size.}
    \label{yieldMax}
\end{figure}
\begin{figure}
	\centering
    \includegraphics[scale=0.9]{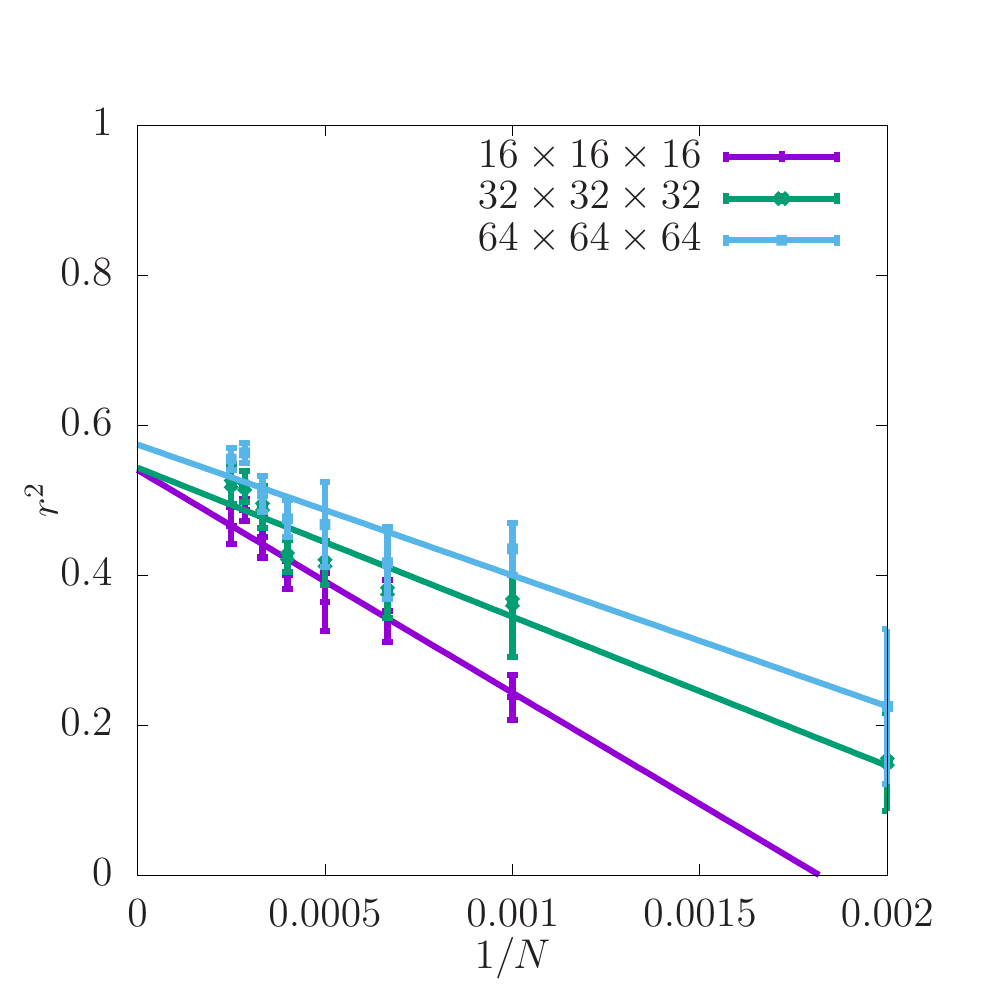}
    \caption{Asymptotic behaviour of $r^2$ of the test set for the maximal stress value along the stress-strain curve obtained for three different resolutions of the CNN input data as a function of $1/N$. The errorbars are standard errors of the mean (SEM).}
    \label{yieldMax_asymptotic}
\end{figure}
\begin{figure}
    \centering
    \includegraphics[scale=0.9]{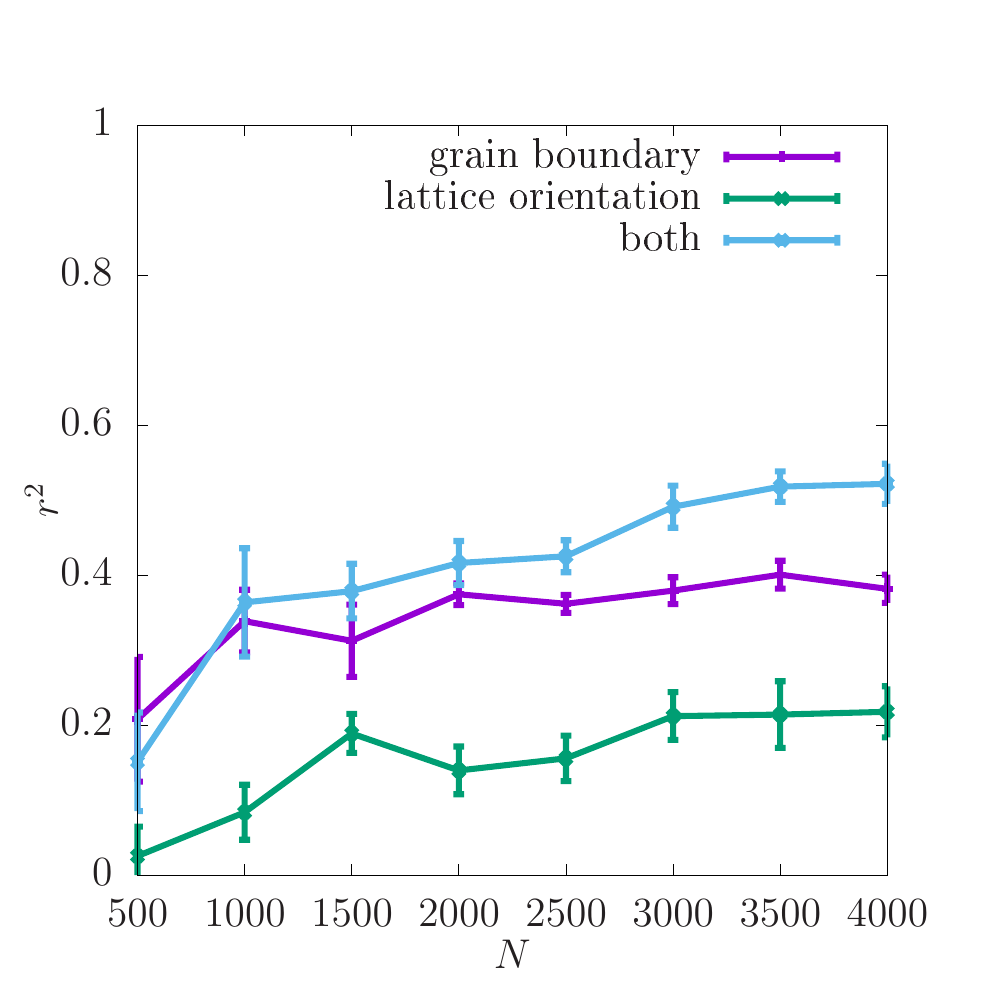}
    \caption{$r^2$ for the maximal stress value along the stress-strain curve obtained for the test set at the resolution $32\times32\times32$ with the descriptors used separately and combined. The errorbars are standard errors of the mean (SEM).}
    \label{yield_descriptors}
\end{figure}
In Fig. \ref{offsetYieldStressResolutions} $r^2$ for the yield stress obtained with the offset value 2\% is shown for three different resolutions as a function of $N$ (with a spacing of 500 configurations). $r^2$ is significantly lower than in the case of the shear modulus and $\delta$ is slightly higher. Similarly to the shear modulus, the predictability of the offset yield stress increases with the resolution and the most significant increase in $r^2$ occurs between $16\times16\times16$ and $32\times32\times32$. Further increasing the resolution to $64\times64\times64$ does not improve the predictability significantly. For all the resolutions $r^2$ increases with $N$ and $\delta$ becomes smaller. According to the asymptotic behaviour shown in Fig. \ref{offsetYieldStressResolutions_asymptotic} the value of $r^2$ of the test set at $N\rightarrow\infty$ is also higher for the resolutions $32\times32\times32$ and $64\times64\times64$ than $16\times16\times16$.

In Fig. \ref{offsetYieldStressDescriptors} the values of $r^2$ for the yield stress obtained with 2\% offset for the separate descriptors are shown. In contrast to the shear modulus for this definition of the yield stress the grain boundary is a more important descriptor than the lattice orientation. The importance of the former increases though with the dataset size $N$.

The coefficient of determination $r^2$ at $32\times32\times32$ for the yield stress for 2\% offset is compared in Fig. \ref{offsetYieldStress} with that obtained for 1\% offset. It can be observed that the predictability is similar for both values of the offset and $r^2$ for the test set reaches 0.6 for the full dataset. $\delta$ is reduced with increasing $N$ in both cases. The asymptotic behaviour shown in Fig. \ref{offsetYieldStress_asymptotic} indicates that the values of $r^2$ are very close to each other for $N\rightarrow\infty$. Again, the corresponding scatter plots of true against predicted values for those two cases are shown in Fig. \ref{scatter_plots}b and c. In accordance with the difference between the $r^2$ values it can be observed that the scatter for yield stress is significantly larger than that for shear modulus.

The coefficient of determination $r^2$ for the training and test set, obtained from the CNNs trained for the maximal stress value is shown in Fig. \ref{yieldMax}, again at three different resolutions of the input data. As for the offset definition, it can be observed that both for the training and the test set the value of $r^2$ increases with $N$. The results look similar for all the resolutions studied. The insets show that $\delta$ decreases with increasing $N$. The corresponding asymptotic behaviour can be seen in Fig. \ref{yieldMax_asymptotic} and the corresponding scatter plot of true against predicted values can be seen in Fig. \ref{scatter_plots}d.
\begin{figure*}
	\centering
    \includegraphics[scale=0.9]{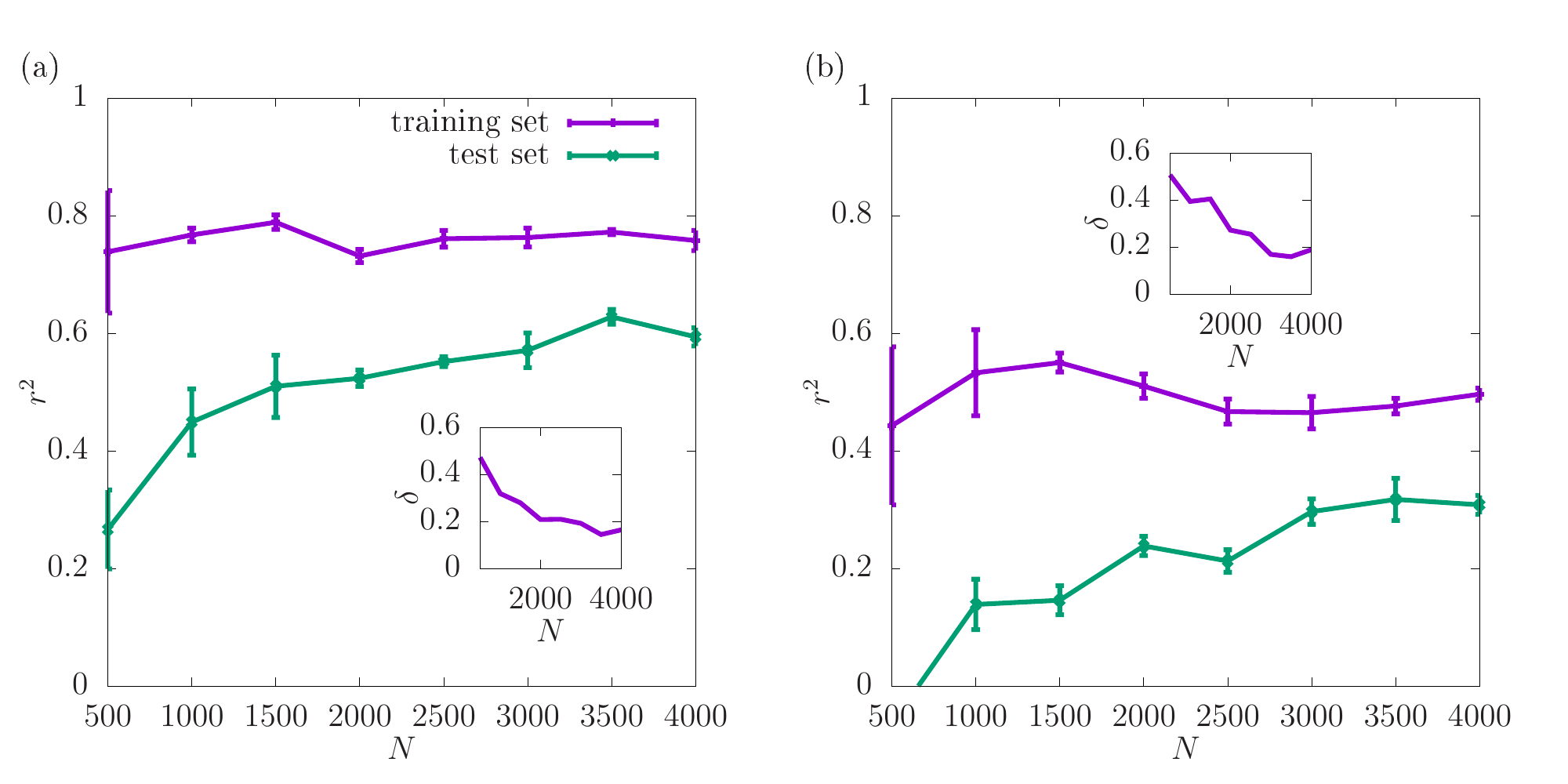}
    \caption{$r^2$ for the stress at the fixed strain value of 0.075 (a) and 0.1 (b), obtained for the resolution $32\times32\times32$ of the CNN input data, shown as a function of $N$. The errorbars are standard errors of the mean (SEM). The insets show $\delta$ as a function of $N$.}
    \label{yieldStressAtFixedStrain}
\end{figure*}
\begin{figure}
	\centering
    \includegraphics[scale=0.9]{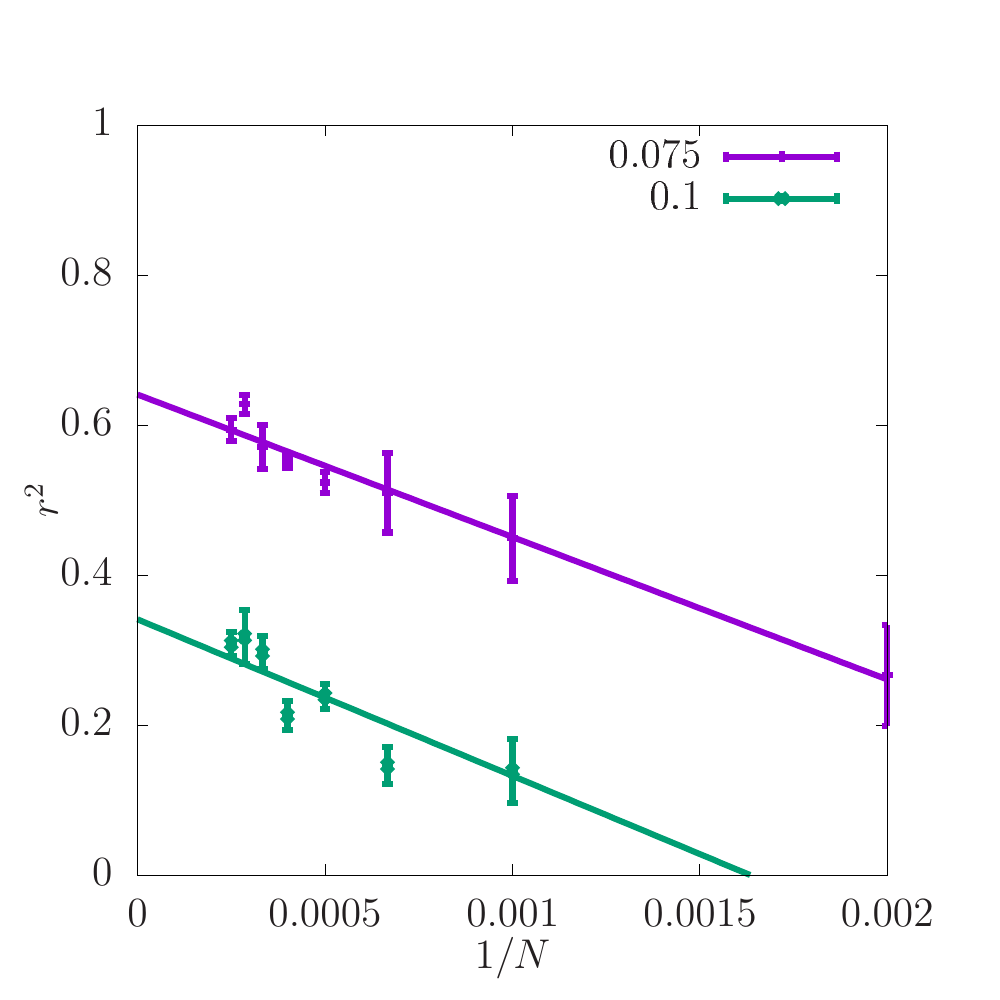}
    \caption{Asymptotic behaviour of $r^2$ of the test set for the stress at the fixed strain value of 0.075 and 0.1, obtained for the resolution $32\times32\times32$ of the CNN input data, shown as a function of $1/N$. The errorbars are standard errors of the mean (SEM).}
    \label{yieldStressAtFixedStrain_asymptotic}
\end{figure}

Again, the values of $r^2$ obtained for the separate descriptors compared to them being used together are shown in Fig. \ref{yield_descriptors}. Also this time it can be seen that the grain boundary is a more important descriptor than the lattice orientation, however, both descriptors provide significant information about the yield stress for all $N$-values considered. The value of $r^2$ for the combination of the descriptors is also significantly higher than for either of the descriptors used separately.

In Fig. \ref{yieldStressAtFixedStrain} the coefficient of determination for the stress value at the fixed strain values of 0.075 and 0.1 is shown. For all $N$-values considered, $r^2$ is larger for the lower strain value. This is expected since the corresponding point of the stress-strain curve lies closer to the elastic part, and, as seen earlier, the elastic properties are predicted much more easily than the plastic properties. The scatter plots of true against predicted values in Fig. \ref{scatter_plots}e and f suggest that the low values of stress at 0.1 strain are overestimated by the CNN, while the high ones are underestimated. Also the asymptotic behaviour shown in Fig. \ref{yieldStressAtFixedStrain_asymptotic} indicates that $r^2$ is much higher for the strain value of 0.075 than 0.1.

\begin{figure}
	\centering
    \includegraphics[scale=0.9]{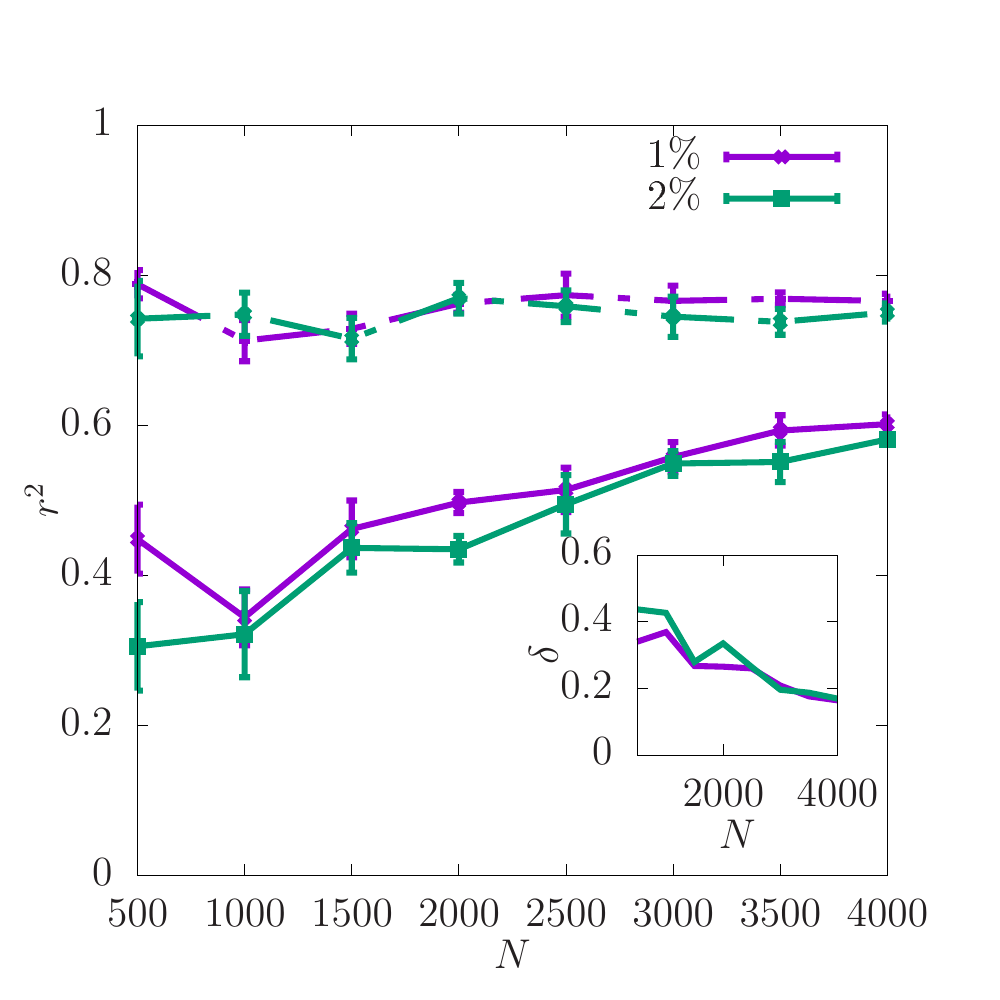}
    \caption{$r^2$ for the yield strain defined with the offset method for the resolution of $32\times32\times32$ of the CNN input data as a function of $N$. The dashed lines show the values of $r^2$ for the training set, and the continuous lines for the test set. The errorbars are standard errors of the mean (SEM). The inset shows $\delta$ as a function of $N$.}
    \label{offsetYieldStrain}
\end{figure}
\begin{figure}
	\centering
    \includegraphics[scale=0.9]{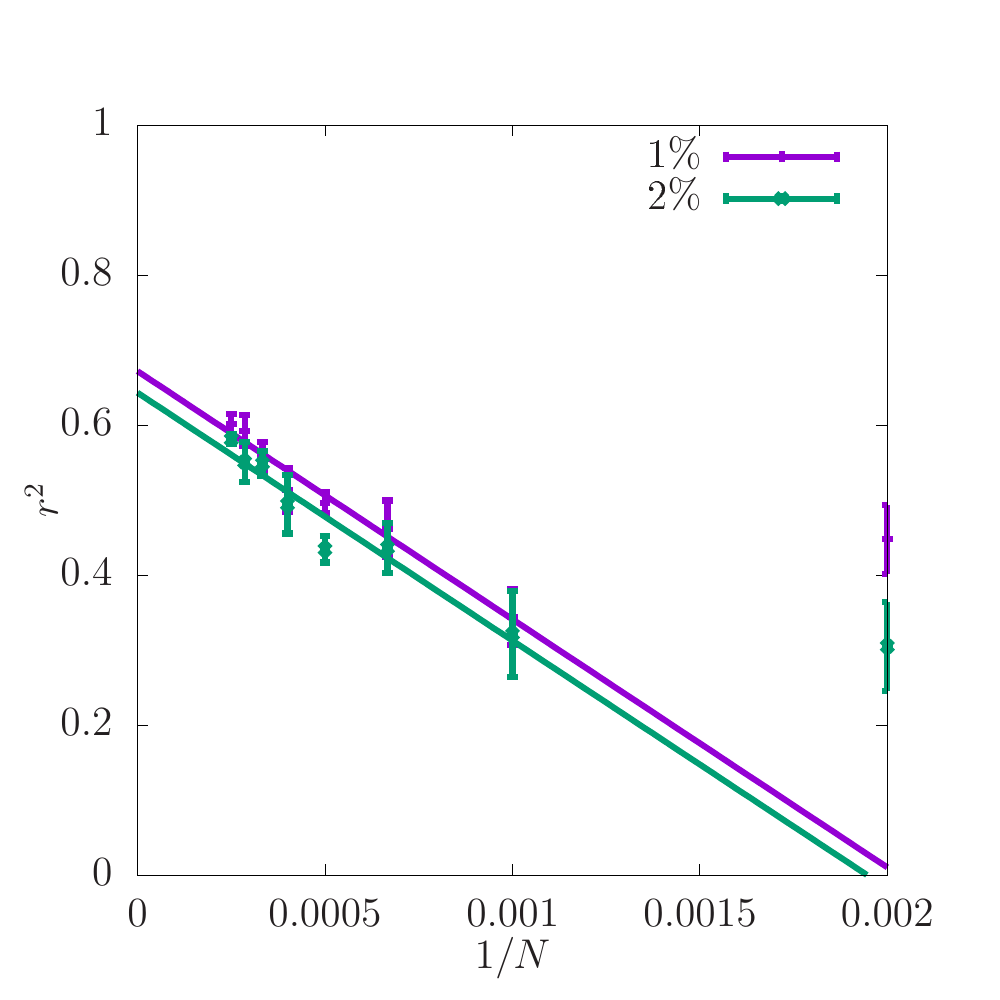}
    \caption{Asymptotic behaviour of $r^2$ of the test set for the yield strain defined with the offset method for the resolution of $32\times32\times32$ of the CNN input data as a function of $1/N$ with the fitting range (0;0.001). The errorbars are standard errors of the mean (SEM).}
    \label{offsetYieldStrain_asymptotic}
\end{figure}
\subsubsection{Yield strain}
In addition to the yield stress, one can also train the CNN to predict the yield strain. Obviously, it only makes sense for the definitions in which the yield point is not determined at the fixed value of strain. Moreover, it turns out that the prediction is very poor for the yield point defined as the maximum stress of the stress-strain curve. This is most likely due to the fact that while the maximum value of stress is to a certain degree determined by the structure of the polycrystal, the strain value at which this maximum is reached is largely random due to the fluctuating character of the stress-strain curve. Therefore, the only definition for which the results of the prediction are presented is the one which employs the offset method. $r^2$ obtained with that method is shown in Fig. \ref{offsetYieldStrain}, again for the offset values of 1\% and 2\%, scatter plots of true against predicted values are shown in Fig. \ref{scatter_plots}g and h and the corresponding asymptotic behaviour is shown in Fig. \ref{offsetYieldStrain_asymptotic}, however the fitting range for the linear function is (0;0.001) due to the points at low $N$ being outliers.

\subsection{Sensitivity to initial conditions}
In all the studied cases the value of $r^2$ is lower than 1, which implies that the predictability is never perfect. Even though $r^2$ for the test set tends to increase with $N$ (and with the resolution) as the gap between the training and the test set closes, for both of those sets $r^2$ seems to approach a certain value below 1. This suggests that there exists a certain limit for the predictability of both the elastic and plastic properties of the small polycrystalline samples we study. This limit might be related to fundamental properties of the system which are manifested as sensitivity of the mechanical response to small variations in the initial conditions of the samples, not properly captured by the descriptors given with a finite resolution.

It has been seen that the predictability for the shear modulus is higher than that for the yield point. Moreover, different definitions of the yield stress resulted in different values of $r^2$. As already shown in Fig. \ref{model} the stress-strain curves exhibit a larger variability near the yield point than in the elastic regime. However, this does not explain directly the difference in the values of the predictability scores between the shear modulus and the yield stress, since according to Eq.~(\ref{coeff_determ}) the predictability is measured as the accuracy of the fit relative to the variance of the quantity that is being predicted. Therefore, {\it a priori} quantities exhibiting a higher variability should not as such be harder to predict.

On the other hand, studying the degree of sensitivity of the deformation dynamics of the system to small perturbations of the initial conditions could give us insight into the limits of predictability. There are several ways in which such sensitivity may limit the predictability score. First of all, the velocities of the particles in the system are initialized randomly. The values of those velocities are not a part of the descriptors fed to the algorithm, however, one can expect that they may influence the details of the dynamics of the system, something that could result in changes in the stress-strain curve from which all the quantities discussed in this work in the context of predictability are extracted. Moreover, the system may be sensitive to the initial choice of the parameters in the Voronoi tessellation, that is the position of the nodes, which specify the shapes of the grains, and the angles for the lattice rotation. Since those features of the system are given to the CNN with a finite resolution, small variations in grain shapes and lattice rotations may not result in any changes to the descriptors. Finally, the computer simulations are performed with finite decimal precision, which leads to inaccuracies in integrating the equations of motion. All those factors may contribute to a limit of the predictability.

In order to study the sensitivity of the system to the initial conditions quantitatively new MD simulations have been carried out using as the initial state sets of configurations in which one of the features described above (random seed, position of the Voronoi nodes and lattice orientation) is varied while the remaining ones are kept the same. To make the results representative of the whole dataset, 15 configurations were picked from the original set and for each of them and for each of the features 49 new simulations were run (making it 50 stress-strain curves including the unperturbed one). To measure the sensitivity quantitatively, one should relate the variance of the system's response, for instance the stress at a given strain or the shear modulus, measured here for different random seeds to the variance of the same response for the whole original set of configurations. The sensitivity $\chi$ (which is taken to be a function of the perturbation $\bm{\alpha}$) can therefore be defined as
\begin{equation}
    \chi(\bm{\alpha})=\frac{\langle\langle(y_i(\bm{\alpha}_j)-\langle y_i(\bm{\alpha}) \rangle)^2\rangle_j\rangle_i}{\langle(y-\langle y \rangle)^2\rangle}.\label{sensitivity_def}
\end{equation}
The denominator is the variance of the quantity $y$ over the whole original set of configurations, while the nominator is the average of the variances of the same quantity determined for the configurations perturbed with the magnitude $\bm{\alpha}$ over the set of the configurations selected for the sensitivity analysis enumerated with the index $i$. The magnitude of perturbation can be written as
\begin{equation}
    \bm{\alpha}=(\Delta x,\Delta y,\Delta z,\Delta\psi,\Delta\theta,\Delta\phi),
\end{equation}
where $\Delta$ denotes the standard deviation, $x$, $y$ and $z$ are positions of the nodes, and $\psi$, $\theta$ and $\phi$ are the Euler angles corresponding to the rotations of the grains. $\bm{\alpha}_j$ is the specific realization of the perturbation with the magnitude $\bm{\alpha}$. $\bm\alpha=(0,0,0,0,0,0)$ implies that changing the random seed is the only perturbation made.

It is easy to see that the definition of sensitivity in Eq. (\ref{sensitivity_def}) is similar to the ratio of the variances in the definition of $r^2$ in Eq. (\ref{coeff_determ}). Both quantities relate the scatter of the values obtained in some procedure to the scatter of the reference values of the system. It can be expected that for sufficiently low $\bm{\alpha}$ in the case when the sensitivity is the only factor limiting the predictability, $\chi=1-r^2$. Since there are always also other predictability-limiting factors (such as limited dataset, convergence of training of the ML algorithm, its complexity, choice of descriptors), in practice one has $\chi<1-r^2$.

\begin{figure}
    \centering
    \includegraphics[scale=0.75]{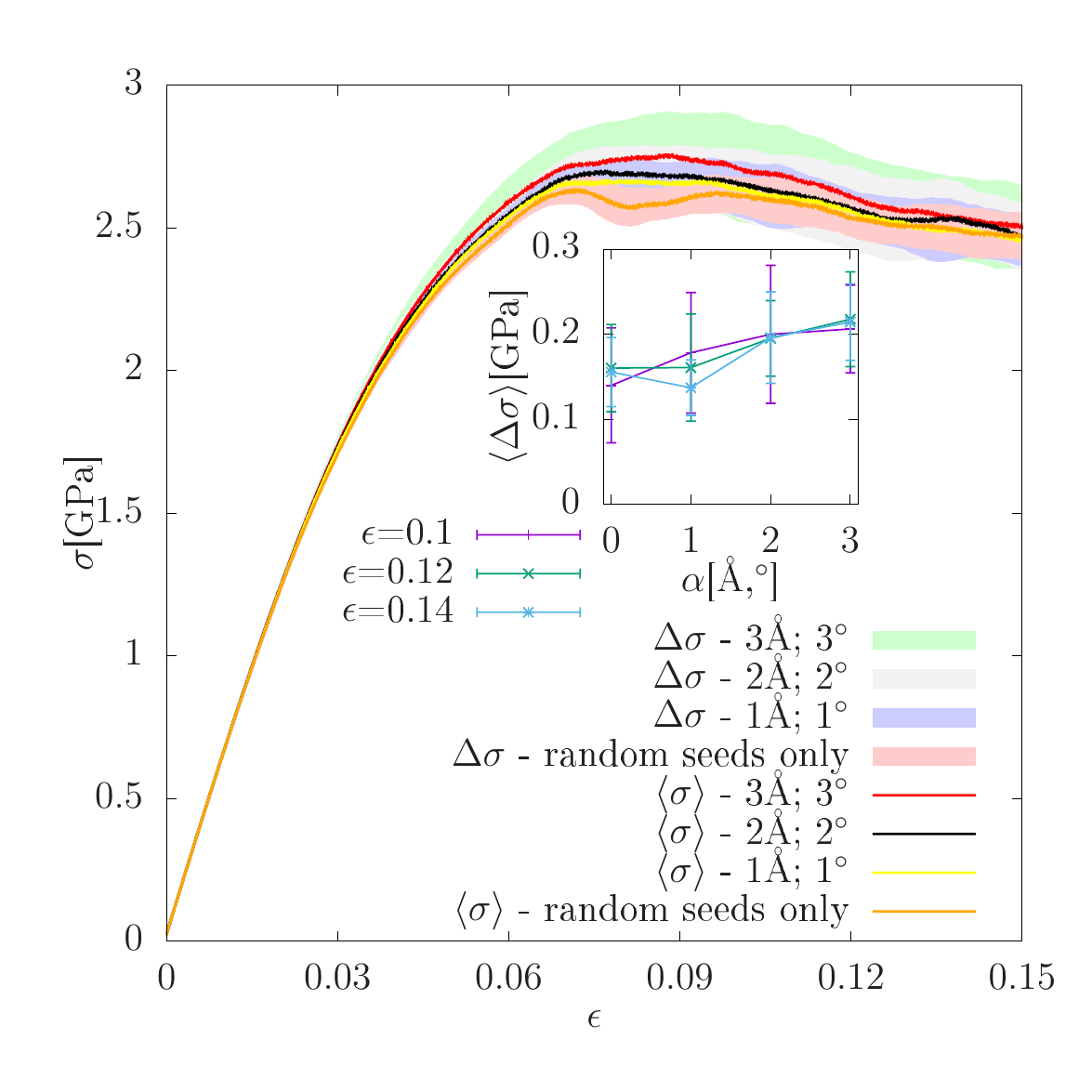}
    \caption{Scatter of stress-strain curves for different magnitudes of the perturbation $\bm{\alpha}$. The inset shows the standard deviation averaged over the all available configurations and the window of the width 0.02 centered at the given strain.}
    \label{scatter_magnitude}
\end{figure}

In order to establish the dependence of the sensitivity on the magnitude of perturbation, MD simulations were performed for configurations with different values of $\bm{\alpha}$. For 15 configurations selected from the original set 49 perturbed configurations with $\bm(3$\AA$,3\degree)$ were generated. For all of them new MD simulations were performed, for each of which a different random seed was used. The same number of simulations was performed for the unperturbed configurations with only changing the random seed. Additionally, three configurations were selected and perturbed by simultaneously displacing the nodes and rotating the lattice orientation with $\bm{\alpha}=(2$\AA$,2\degree)$ in one set, and $\bm{\alpha}=(1$\AA$,1\degree)$ in the other set. For one chosen configuration the corresponding averaged stress-strain curves together with their scatter are shown in Fig. \ref{scatter_magnitude} for different magnitudes of perturbation $\bm{\alpha}$. As can be seen there the standard deviation of the stress response increases with $\bm{\alpha}$, however, in the range of $\bm{\alpha}$ shown in the inset it differs at most by a factor of 2. Therefore, because of the weak dependence on $\bm{\alpha}$, in the discussion below $\chi(3$\AA$,3\degree)$ will be used.

Below the sensitivity of the system to perturbations of different initial conditions is discussed. The extracted values of $\chi$ for different types of perturbation are collected in Table \ref{sensitivity_table}, where they are compared with the corresponding values of $r^2$ obtained for the full data set at the resolution $32\times32\times32$.

\subsubsection{Random seeds}
\begin{figure*}
	\centering
    \includegraphics[scale=0.6]{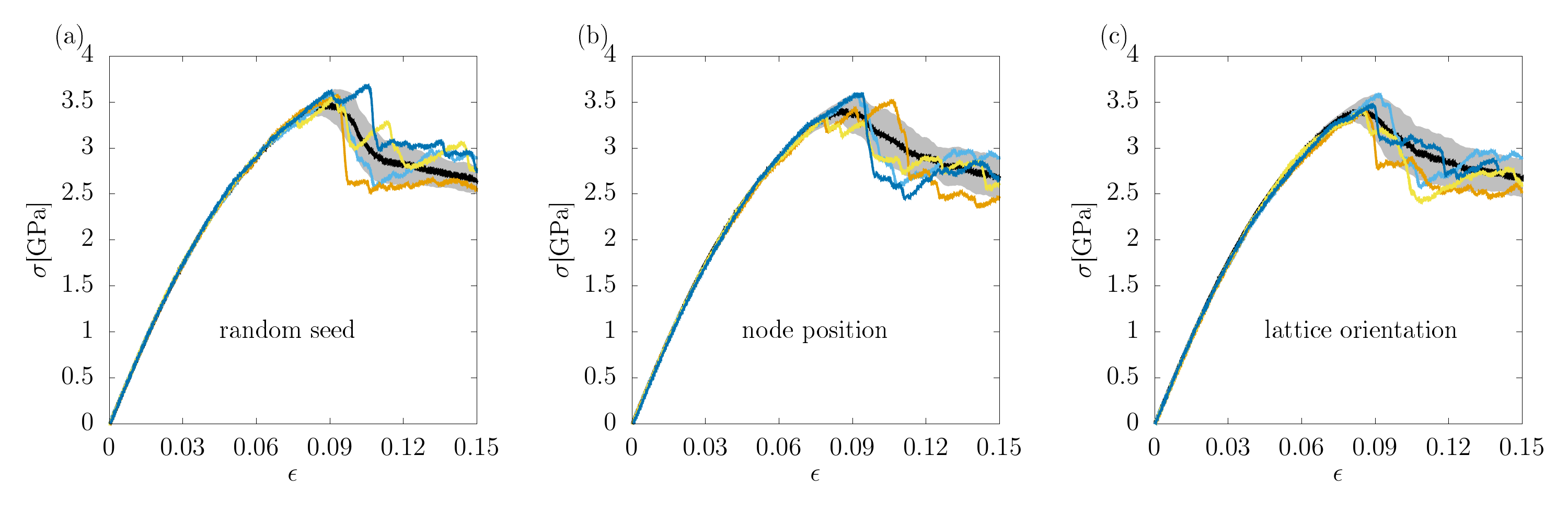}
    \caption{Stress-strain curves generated for an individual example initial polycrystalline microstructure which is perturbed in different ways, to assess the sensitivity of the response to small perturbations of the initial state. (a) Perturbations in the form of different random seeds used to initialize the atom velocities, (b) perturbations of the Voronoi nodes with $\bm{\alpha}=(3$\AA$,0)$, and (c) perturbations of the angular orientation of the grains with $\bm{\alpha}=(0,3\degree)$. The thick black lines are the averages over 50 stress-strain curves, each obtained for a different perturbation of the initial state. The gray areas represent the standard deviation of the stress at a given strain.}
    \label{stressStrainPerturbation}
\end{figure*}

\begin{figure}
    \centering
    \includegraphics[scale=0.9]{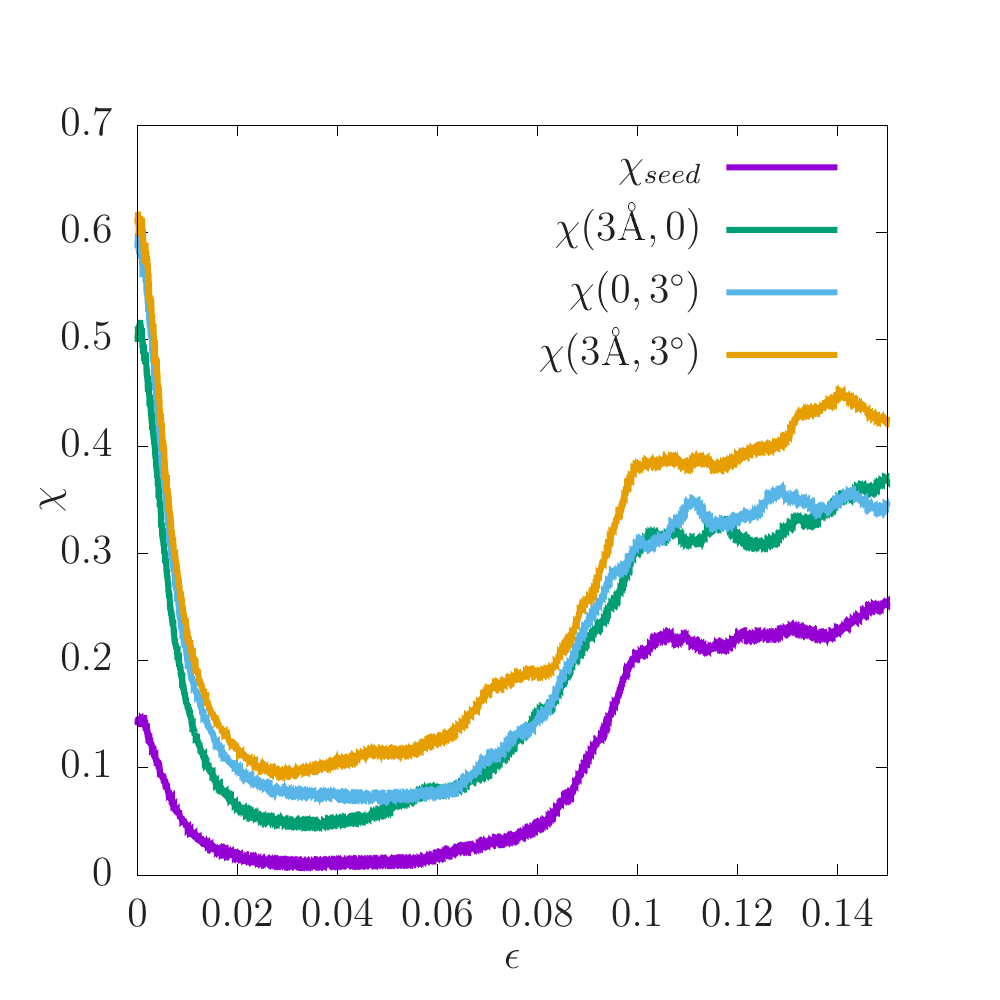}
    \caption{$\chi$ of the stress value at the given strain measured for the choice of the random seed initializing the velocities of the atoms, perturbation of the position of the nodes in the Voronoi tessellation, perturbation of the lattice rotation of the grains, and all of those perturbations combined.}
    \label{sensitivity}
\end{figure}

First, simulations starting with the same initial polycrystalline configuration but different random seeds initializing the velocities of the particles at the start of the equilibration phase ($\bm{\alpha}=(0,0)$) have been carried out. 15 different configurations were used, for each of which 50 simulations with different random seeds have been performed. The stress-strain curves for several such random seeds for one of the configurations are shown in Fig. \ref{stressStrainPerturbation}a. As can be seen there, the curves are similar in the elastic regime, resulting in similar values of the shear modulus, however, in the plastic regime there is a large variability of the stress. In Fig. \ref{sensitivity} the sensitivity $\chi_{seed}=\chi(0,0)$ of the stress at the given strain value to the initial choice of the atom velocities is shown. It can be observed that its value is relatively high for very low values of strain, which is most likely due to the thermal fluctuations of the stress being at the beginning of the deformation larger than its average value. Around the strain value of 0.02 $\chi_{seed}$ drops to a very low value and stays there until the yield occurs, that is, around 0.08 strain. Above that $\chi_{seed}$ reaches values slightly over 0.2. This result can be used to explain the difference in the values of $r^2$ for the stress at the fixed strain value. For the strain value of 0.075 $r^2$ is significantly higher than for 0.1, and also an increase in $\chi_{seed}$ occurs between those two values.

For the the shear modulus, yield strain and the other definitions of the yield stress one can perform the same analysis in an analogous way, namely by determining the variance of the chosen value for the random seeds, dividing it by the variance of the same value for the original set and averaging over the 15 configurations, for which the sensitivity analysis has been performed. They are all shown in Table \ref{sensitivity_table}. The lowest $\chi_{seed}$ is found for the shear modulus. It is also in agreement with the fact that $r^2$ for the shear modulus is the highest for all the quantities studied, approaching the value of 0.9 for the full dataset and the highest resolution. On the other hand, the highest $\chi_{seed}$ is exhibited by the stress at 0.1 strain, which also has the lowest $r^2$. $\chi_{seed}$ for the maximal value of the stress is higher than for the fixed strain value of 0.075 but lower than that at the strain value of 0.1. Finally, the values of $\chi_{seed}$ for the yield point determined with the offset method are relatively low, which is in agreement with their high values of $r^2$.

\begin{table*}
    \begin{tabular}{|l|l|l|l|l|l|l|l|}
        \hline
        quantity & $\chi_{seed}$ & $\chi(3$\AA$,0)$ & $\chi(0,3\degree)$ & $\chi(3$\AA$,3\degree)$ & $1-\chi(3$\AA$,3\degree)$ & $r_{asymptotic}^{2}$ \\ \hline
        shear modulus & 0.005 & 0.012 & 0.043 & 0.054 & 0.946 & 0.889 $\pm$ 0.011\\ \hline
        1\% offset yield stress & 0.025 & 0.097 & 0.102 & 0.164 & 0.836 & 0.662 $\pm$ 0.012\\ \hline
        2\% offset yield stress & 0.028 & 0.107 & 0.098 & 0.16 & 0.84 & 0.659 $\pm$ 0.012 \\ \hline
        maximal stress & 0.05 & 0.131 & 0.121 & 0.176 & 0.824 & 0.543 $\pm$ 0.017\\ \hline
        stress at 0.075 strain & 0.034 & 0.121 & 0.128 & 0.183 & 0.817 & 0.641 $\pm$ 0.011 \\ \hline
        stress at 0.1 strain & 0.205 & 0.302 & 0.31 & 0.38 & 0.62 & 0.341 $\pm$ 0.02 \\ \hline
        1\% offset yield strain & 0.038 & 0.111 & 0.12 & 0.179 & 0.821 & 0.672 $\pm$ 0.014 \\ \hline
        2\% offset yield strain & 0.039 & 0.116 & 0.111 & 0.169 & 0.831 & 0.643 $\pm$ 0.021\\ \hline
    \end{tabular}
    \caption{Values of sensitivity $\chi$ for different quantities compared with their asymptotic predictability $r_{asymptotic}^{2}$ obtained at the resolution of $32\times32\times32$. Also values of $1-\chi(3$\AA$,3\degree)$, i.e., the theoretical maxima of the $r_{asymptotic}^2$'s, are shown. The errors for $r_{asymptotic}^{2}$ are asymptotic standard errors.}
    \label{sensitivity_table}
\end{table*}

\subsubsection{Nodes in Voronoi tessellation}
In order to study the sensitivity of the system to the positions of the nodes in the Voronoi tessellation, for each of the 15 chosen configurations additional 49 configurations were generated by displacing the nodes randomly according to a Gaussian distribution with the standard deviation equal to 3\AA, which corresponds to $\bm{\alpha}=(3$\AA$,0)$. This procedure leads to configurations similar to the original ones with slightly different grain structures but the same lattice orientation within the grains. The resulting stress-strain curves are shown in Fig. \ref{stressStrainPerturbation}b. $\chi(3$\AA$,0)$ for the stress value at a given strain, determined in an analogous way to the case of different random seeds, is shown in Fig. \ref{sensitivity}. It can be seen there that $\chi(3$\AA$,0)$ is significantly higher than $\chi_{seed}$. In particular, its magnitude at the strain values of 0.075 and 0.1, used for the prediction of the stress, is also higher in this case.

In the analogous way as in the case of different random seeds, $\chi(3$\AA$,0)$ for other quantities of the system was determined. It can be seen in Table \ref{sensitivity_table} that the values of $\chi(3$\AA$,0)$ are generally higher than the corresponding values of $\chi_{seed}$. However, the relations between its values for different quantities are similar. The minimal and the maximal $\chi(3$\AA$,0)$ is again exhibited by the shear modulus and the stress at 0.1 strain, respectively.
\subsubsection{Lattice orientation of grains}
Next, the sensitivity of the system to the initial lattice orientation of the grains inside the polycrystal was studied. This time new configurations were generated with the fixed position of the nodes in the Voronoi tessellation and Euler angles perturbed according to a Gaussian distribution with the standard deviation equal to 3$\degree$, that is $\bm{\alpha}=(0,3\degree)$. The stress-strain curves corresponding to that perturbation are shown in Fig. \ref{stressStrainPerturbation}c. Again, the magnitude of $\chi(0,3\degree)$ at different strain values is shown in Fig. \ref{sensitivity}. It seems higher in the elastic regime compared to the previous measures of $\chi$. Furthermore, as shown in Table \ref{sensitivity_table}, $\chi(0,3\degree)$ for the shear modulus is also higher in this case, which is related to the previously mentioned observation that the elastic properties of the sample are controlled mainly by the lattice orientation, which is the perturbed property here. On the other hand, $\chi(0,3\degree)$ in the plastic part of stress-strain curve is comparable to $\chi(3$\AA$,0)$ and for some of the quantities it is actually slightly smaller. Unlike the previous measures of $\chi$, $\chi(0,3\degree)$ for the yield point determined with the offset method is lower for the higher offset than for the lower one. Again, this is related to the fact that the lattice orientation of the grains has a larger impact on the elastic properties of the sample than on the plastic ones.
\subsubsection{Total sensitivity}
All the contributions to the sensitivity discussed above (random seed, position of nodes and lattice orientation) contribute to the total sensitivity of the system to the initial conditions. However, one cannot expect that the measure of the total sensitivity $\chi_{total}=\chi(3$\AA$,3\degree)$ is simply a sum of all those contributions. Therefore, additional MD simulations were carried out. They were performed in an analogous way as before but instead of changing only one of the initial parameters discussed above, all three of them were varied simultaneously. The value of $\chi(3$\AA$,3\degree)$ as the function of strain is also shown in Fig. \ref{sensitivity}. It can be seen that it is always higher than $\chi$ measured with respect to change of any of the features separately.

\begin{figure}
    \centering
    \includegraphics[scale=0.9]{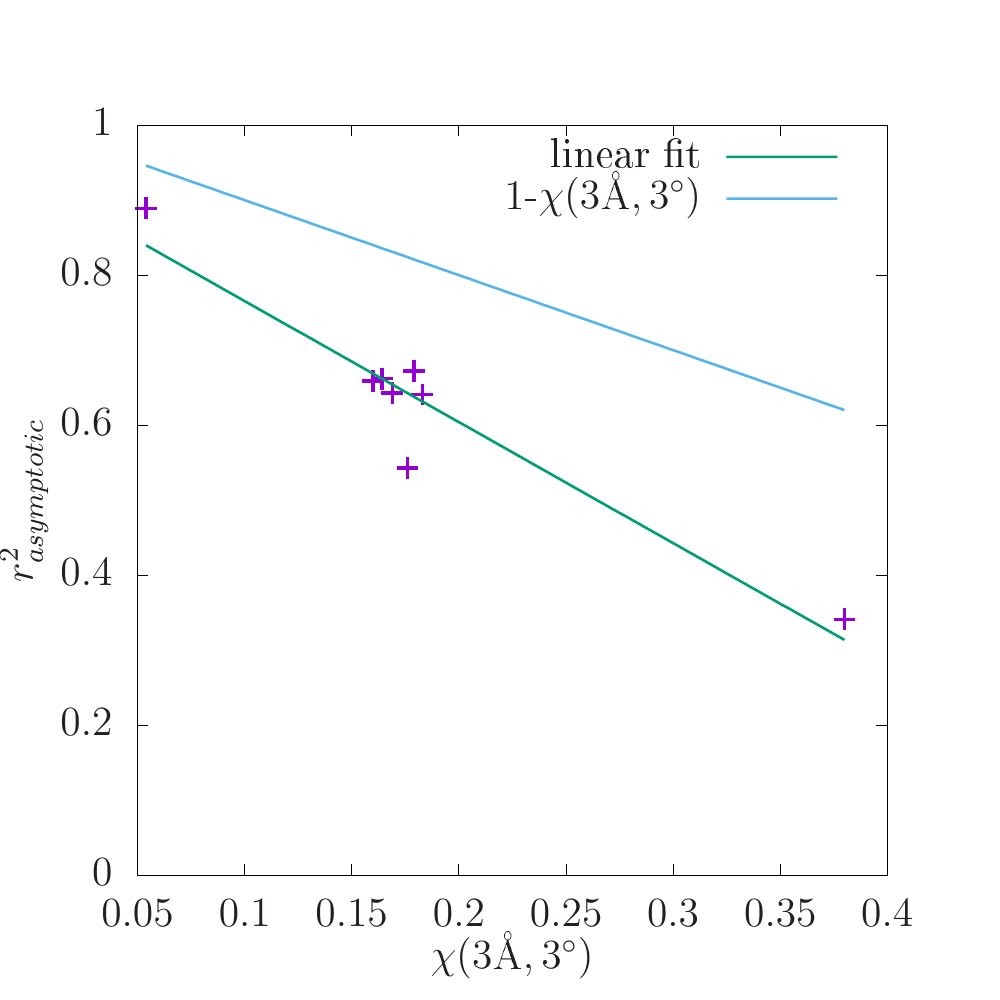}
    \caption{Values of $r_{asymptotic}^2$ of different quantities studied in this work at the resolution of $32\times32\times32$ plotted against their total sensitivity $\chi(3$\AA$,3\degree)$ to the initial conditions of the sample. The green line is a linear fit to the data and the blue line represents the maximal possible $r_{asymptotic}^2$ for the given $\chi(3$\AA$,3\degree)$, i.e., $1-\chi(3$\AA$,3\degree)$.}
    \label{sensitivityPredictability}
\end{figure}

Values of $\chi(3$\AA$,3\degree)$ for all the quantities studied were determined and the results are again collected in Table \ref{sensitivity_table}. They can be compared to the corresponding values of asymptotic predictability $r_{asymptotic}^2$ determined as the intercept in the $1/N$ plots for the resolution of descriptors $32\times32\times32$ and it can be seen that the quantities that are more sensitive to the initial conditions of the system (larger $\chi(3$\AA$,3\degree)$) tend to have smaller $r^2$. The values of $r_{asymptotic}^2$ are plotted against those of $\chi(3$\AA$,3\degree)$ in Fig. \ref{sensitivityPredictability}, where a linear correlation between them can be observed. Additionally, the maximal value of predictability at the given $r_{asymptotic}^2$ equal to $1-\chi(3$\AA$,3\degree)$ is shown in the plot as the blue line. For all the quantities the actual $r_{asymptotic}^2$ lies below that line, which means that the condition $r_{asymptotic}^2 \le 1-\chi(3$\AA$,3\degree)$ is always satisfied.

Some possible additional factors besides $\chi$ that may limit the predictability have been already mentioned, however, it can also be noted that the difference between the maximal ($1-\chi(3$\AA$,3\degree)$) and actual predictability ($r^2$) differs between the quantities. It is the lowest for the shear modulus and the highest for the stress at 0.1 strain. Those two quantities exhibit also the highest and the lowest $r^2$, respectively. Therefore, it seems that those other predictability-limiting factors have different contributions for different quantities. Generally, it can be expected that the elastic properties, such as shear modulus, are relatively easy to predict because their measurement requires only small deformation of the sample, which corresponds to a short time evolution of the system. Moreover, it has been shown that the value of the shear modulus is mostly determined by the lattice orientation of the individual grains. Therefore, it can be expected that the relation between the descriptors and the predicted value is relatively simple. On the other hand, the plastic properties of the sample, such as the value of stress at the strain of 0.1, can be more difficult to predict because they occur further on the stress-strain curve, such that the system might have partially lost its memory of its initial state.
\section{Conclusions}
\label{section:conclusions}
In this work, predictability (measured as the coefficient of determination $r^2$) of the elastic and plastic properties of nano-sized shear-deformed iron polycrystals has been determined using ML methods. While the shear modulus can be easily defined as the slope of the stress-strain curve for small strains, defining the yield stress is not so straightforward and therefore several definitions have been considered. For all those quantities studied it was found that the predictability is increasing with the dataset size $N$ and the spatial resolution of the chosen microstructural descriptors. However, it always seems to reach a certain value below 1, which implies that there exists a certain fundamental limit of deformation predictability of small polycrystals. Moreover, the predictability obtained by the CNN was found to be higher for the shear modulus than for the yield point, independently of the exact definition of the latter. The reasons for that difference and for the limit of predictability have been explored by measuring the sensitivity of the system studied to small perturbations of its initial state.

This sensitivity of the system has been measured by varying the random seed initializing the velocities in the MD simulation, position of the nodes in the Voronoi tessellation and lattice orientations of the grains inside the initial polycrystal configuration. It has been found that in accordance with the differences in the predictability the plastic properties of the system exhibit a larger sensitivity to the initial state than the elastic properties. In general, the sensitivity can be thought of as a measure of the amount of information that is not available to the ML algorithm. Since at any finite temperature the system constantly fluctuates and its descriptors are extracted from the equilibrated configuration at some timestep, the magnitude of the fluctuations of the position of the atoms and their velocities are unknown to the CNN. Moreover, a voxelized representation of the initial microstructure with any finite resolution tends to hide small differences in the initial microstructure between samples. Therefore, given that the system exhibits sensitivity to small perturbations of the initial microstructure, two configurations with identical descriptors may result in different time evolution and, as studied here, different stress-strain curves.

Our study thus provides important insights into the fundamental limits of deformation predictability, and those insights are expected to apply more generally to predicting the time evolution of complex physical systems. Even if the dynamics of a complex system, such as deformation of the polycrystals studied here, is governed by deterministic equations of motion, its predictability may still be limited due to the incomplete information about the initial state and other factors such as random thermal fluctuations. Even though the study presented here is purely computational, the conclusions drawn from it could also be extended to experiments. The accuracy at which the structure of the sample can be determined by measurements is always finite and because of the constant thermal fluctuations essentially no information is available about the velocities of individual particles. Overall, the analysis presented here concerning the role of sensitivity of the system to small perturbations of its initial configuration in limiting the predictability of the system's future evolution could be applied in a wide range of contexts where one aims at predicting the behaviour of a complex system.
\section{Methods}
\label{section:methods}
\subsection{Generation of polycrystals}
\label{subsection:methodsPolycrystals}
The tools used for generation of the polycrystalline samples are Atomsk \cite{hirel2015atomsk} and Nanocrystal generator \cite{nanogen}; the latter program has been developed in our research group. Both programs implement the Voronoi tessellation \cite{brostow1978construction,finney1979procedure}, which is a method to partition the three-dimensional space into a set of polyhedra, which are here taken to represent the individual grains of the polycrystal. Voronoi tessellation is a common way of generating polycrystals, which exists in many variants \cite{falco2017generation}. It is fully defined by specifying positions of a certain number of points, called seeds, in the space. For each of those seeds there is a corresponding region called a Voronoi cell which contains all the points that are closer to that given seed than to any other seed. Those Voronoi cells are subsequently filled with atoms arranged in the chosen lattice structure and with the specified crystallographic orientation to represent grains of the polycrystal. In all three directions periodic boundary conditions are implemented.

In this work both the positions of the seeds and the Euler angles for the crystallographic orientation of the individual grains are chosen randomly for each polycrystal. The uniform distribution of the Euler angles results, however, in a non-uniform distribution of crystal orientations. Therefore, the polycrystalline samples studied in this work are in fact textured polycrystals with the rotation of the grains biased towards the poles along the z-direction \cite{perez2013uniform}.

Some sets of parameters in the Voronoi tessellation resulted in configurations that were not stable due to the distances between the atoms across the grain boundary being too small. Such configurations were removed and replaced with new ones that did not result in this problem.

\subsection{Molecular dynamics simulations}
\label{subsection:methodsMD}
As the interatomic potential for the MD simulations the embedded atom model (EAM) potential for Fe \cite{mendelev2003development} is used. During a single MD run first the potential energy of the configuration is minimized by letting the atoms relax and then the system is equilibrated at the constant temperature of 300 K using the Nose-Hoover thermostat, and zero pressure in the NPT ensemble. Finally the shear deformation of the sample is performed in the NPH ensemble in $xy$ plane under constant strain rate, which is done by tilting the simulation box. Due to the choice of the ensemble the temperature is allowed to change, that is, no thermostat is used. That would assure that the equations of motion are indeed deterministic, which would be analogous, e.g., to discrete dislocation dynamics (DDD) simulations where no thermal noise is present. Notice however that randomness is included via different randomly chosen initial velocities of the atoms for each sample. During the whole deformation run the instantaneous $xy$ component of the pressure tensor is stored as the function of time.

The timestep used in all the MD simulations is 1 fs. After the equilibration run that lasts 1 ns, the MD shear deformation run is performed at the constant strain rate of 3$\cdot$10$^{8}$/s until the strain reaches the value of 0.15. The $xy$ component of the pressure tensor is stored every 50 timesteps.

\subsection{Descriptors}
\label{subsection:methodsDescriptors}
Two different three-dimensional fields are extracted from the equilibrated configuration of the polycrystals and later used as the input descriptors for the ML algorithm. One of them is the local orientation of the lattice given by the quaternion representation of the rotations in three-dimensional space. A quaternion consists of four components and can be written as $\mathbf{q}=\cos(\Theta/2)+\mathbf{u}\sin(\Theta/2)$, where $\mathbf{u}=(u_x,u_y,u_z)$ is a unit vector in three-dimensional space and $\Theta$ is the angle of rotation around that vector. Since this descriptor gives the information about the local crystallographic orientation, it also encodes the misorientation angles between the grains, which might be relevant for the quantities predicted in the work. The other descriptor is the local density of atoms at the grain boundary, which is identified by removing all the atoms belonging to the bcc structure of the grains. Another descriptor that was tried was the local potential energy. While it was expected that it could contain some important information related to specific misorientation angles of the grain boundaries, as is the case in the coincidence site lattice model, it was found that including that additional descriptor does not improve the predictability score in any way.

Both descriptors used in the work, illustrated for an example configuration in the left-hand part of Fig. \ref{cnn}, are extracted by the OVITO software \cite{stukowski2009visualization}, which provides features able to identify the local structure type (common neighbor analysis \cite{honeycutt1987molecular}) and the crystallographic orientation (polyhedral template matching \cite{larsen2016robust}). The descriptors are used for predicting the shear modulus and the yield stress by means of a CNN.

As was shown earlier, by training a CNN with each of those descriptors separately it was found that for predicting the shear modulus the orientation of the grains is more important, while for predicting the yield point the grain boundary is a more useful descriptor. However, the predictability is always the highest when both the descriptors are used. Therefore the descriptors are combined into five different arrays (four for the lattice orientation and one for the grain boundary).

\subsection{Convolutional neural networks}
\label{subsection:methodsCNN}
A CNN is a ML algorithm which takes as the input a pixelized image of the system and processes it through a set of filters in convolutional and pooling layers. Since the system studied here is three-dimensional, the input arrays consist of voxels, which are equivalents of pixels in three dimensions. In this work a CNN is trained to predict the characteristic features of the stress-strain curves mentioned above: shear modulus and yield stress according to its various definitions.

The input arrays are prepared in several different resolutions, which represent the accuracy in which the field extracted from the given configuration is sampled. The highest one is $64\times64\times64$ because for that resolution the number of voxels is of the same order of magnitude as the number of atoms. The lower resolutions used are $16\times16\times16$ and $32\times32\times32$. The array for the local lattice orientation was created by scanning over all the atoms in the system and assigning their quaternion values to that element of the array whose centre of the corresponding cell is closest to the given atom. If later another atom was found to be even closer to the centre of that cell, the quaternion value assigned to that cell was replaced with the new one. On the other hand, the array for the local density of atoms at the grain boundary was prepared by assigning the number of atoms identified as belonging to the boundary within the given cell to the corresponding element of the array.

The data from the input arrays is passed to and subsequently processed by the convolutional, periodic padding and pooling layers included in the architecture of the CNN. The convolutional layers contain 8 filters. The size of the kernel is $3\times3\times3$ and the stride length is 1 in each direction. The role of the periodic padding layer is to keep the size of the array to be the same as before the convolutional layer by extending it periodically by 1 at each of the edges. The maximum pooling layers reduce each of the spatial dimensions of the data by half. It is done by dividing the input into cubes of dimensions $2\times2\times2$ and selecting the maximal value from each of them. A sequence of a convolutional and pooling layer is repeated as many times as required to reduce the size of the array to the dimension $1\times1\times1\times8$. Therefore, the total number of those layers depends on the input resolution. The activation function in the first convolutional layer is sigmoid, while in all the following ones rectifier functions are used. It was found that this choice of activation functions increases the performance of the training. Additionally, another channel with fewer convolutional, periodic padding and pooling layers but larger filters in the latter, which leads to faster size reduction of the arrays, is added in parallel to the main one. The output of both channels is finally flattened and concatenated giving the linear array of the size 16, which is further processed by a fully connected layer giving a single number representing either the shear modulus or the yield stress as the output.

For the training of the CNN the Adam optimizer is utilized with the learning rate 5$\cdot$10$^{-5}$. The L2 regularization is applied to all the convolutional layers with the parameter $\lambda$=0.001.

To test the convergence of the CNN training, the procedure is performed for different sizes of the dataset, starting from 500 or 1000 configurations and increasing it successively by a certain number until the full dataset is covered. For each of the quantities studied five different CNNs are trained for different random seeds representing different splits of the dataset into the training, test and validation set in the ratio 80:10:10\%. The purpose of the validation set is to interrupt training at the epoch at which the corresponding value of the loss function for that set reaches its minimum value. In this work, the early stopping criterion was used, with the patience of 500 epochs, after which the training is interrupted if there has been no decrease in the loss function of the validation set. The final parameters of the CNN are chosen from that epoch at which the loss function of the validation set has the minimum value.

\section*{Data availability}
\noindent
All data included in this work are available from the corresponding author (MM) upon request.

\bibliography{manuscript}

\section*{Acknowledgments}
\noindent
The authors acknowledge the support of the Academy of Finland via the Academy Project COPLAST (project no. 322405). The authors would like to thank Henri Salmenjoki for interesting discussions on machine learning.
\section*{Author contributions}
\noindent
MM generated the training database, performed the molecular dynamics simulations, trained the machine learning algorithms and wrote the initial version of the manuscript. LL designed and supervised the project. Both authors contributed to the writing of the manuscript.
\section*{Competing interests}
\noindent
The authors declare no competing interests.
\end{document}